\documentclass[sn-nature]{sn-jnl}

\usepackage{graphicx}%
\usepackage{multirow}%
\usepackage{amsmath,amssymb,amsfonts}%
\usepackage{amsthm}%
\usepackage{mathrsfs}%
\usepackage[title]{appendix}%
\usepackage{xcolor}%
\usepackage{textcomp}%
\usepackage{manyfoot}%
\usepackage{booktabs}%
\usepackage{algorithm}%
\usepackage{algorithmicx}%
\usepackage{algpseudocode}%
\usepackage{listings}%
\usepackage{xcolor}

\usepackage{longtable}
\usepackage{lscape}
\usepackage{array}
\usepackage{hyperref}
\newcolumntype{C}[1]{>{\centering\arraybackslash}p{#1}}

\usepackage{color,soul}

\newcommand{\nocontentsline}[3]{}
\newcommand{\tocless}[2]{\bgroup\let\addcontentsline=\nocontentsline#1{#2}\egroup}

\begin{document}

\title[Article Title]{Dataset Artefacts are the Hidden Drivers of the Declining Disruptiveness in Science}


\author*[1]{\fnm{Vincent} \sur{Holst}}\email{vincent.thorge.holst@vub.be}

\author[1]{\fnm{Andres} \sur{Algaba}}\email{andres.algaba@vub.be}

\author[1]{\fnm{Floriano} \sur{Tori}}\email{floriano.tori@vub.be}

\author[2]{\fnm{Sylvia} \sur{Wenmackers}}
\email{sylvia.wenmackers@kuleuven.be}

\author*[1,3]{\fnm{Vincent} \sur{Ginis}}\email{vincent.ginis@vub.be}

\affil[1]{\orgdiv{Data Analytics Laboratory}, \orgname{Vrije Universiteit Brussel}, \orgaddress{\street{Pleinlaan 2}, \city{Brussels}, \postcode{1050}, \country{Belgium}}}

\affil[2]{\orgdiv{Centre for Logic and Philosophy of Science (CLPS)}, \orgname{KU Leuven}, \orgaddress{\street{K.~Mercierplein, 2 – box 3200}, \city{Leuven}, \postcode{3000}, \country{Belgium}}}

\affil[3]{\orgdiv{School of Engineering and Applied Sciences}, \orgname{Harvard University}, \orgaddress{\street{9 Oxford Street}, \city{Boston}, \postcode{MA 02134}, \state{Massachusetts}, \country{USA}}}





\maketitle
Park et al.~\citep{park2023papers} reported a decline in the disruptiveness of scientific and technological knowledge over time. Their main finding is based on the computation of CD indices, a measure of disruption in citation networks \citep{funk2017dynamic}, across almost 45~million papers and 3.9~million patents. 
Due to a factual plotting mistake, database entries with zero references were omitted in the CD index distributions, hiding a large number of outliers with a maximum CD index of one, while keeping them in the analysis~\citep{park2023papers}. Our reanalysis shows that the reported decline in disruptiveness can be attributed to a relative decline of these database entries with zero references. Notably, this was not caught by the robustness checks included in the manuscript. The regression adjustment fails to control for the hidden outliers as they correspond to a discontinuity in the CD index. Proper evaluation of the Monte-Carlo simulations reveals that, because of the preservation of the hidden outliers, even random citation behaviour replicates the observed decline in disruptiveness.
Finally, while these papers and patents with supposedly zero references are the hidden drivers of the reported decline, their source documents predominantly do make references, exposing them as pure dataset artefacts.

\begin{figure}[ht!]%
\centering
\includegraphics[width=1.0\textwidth]{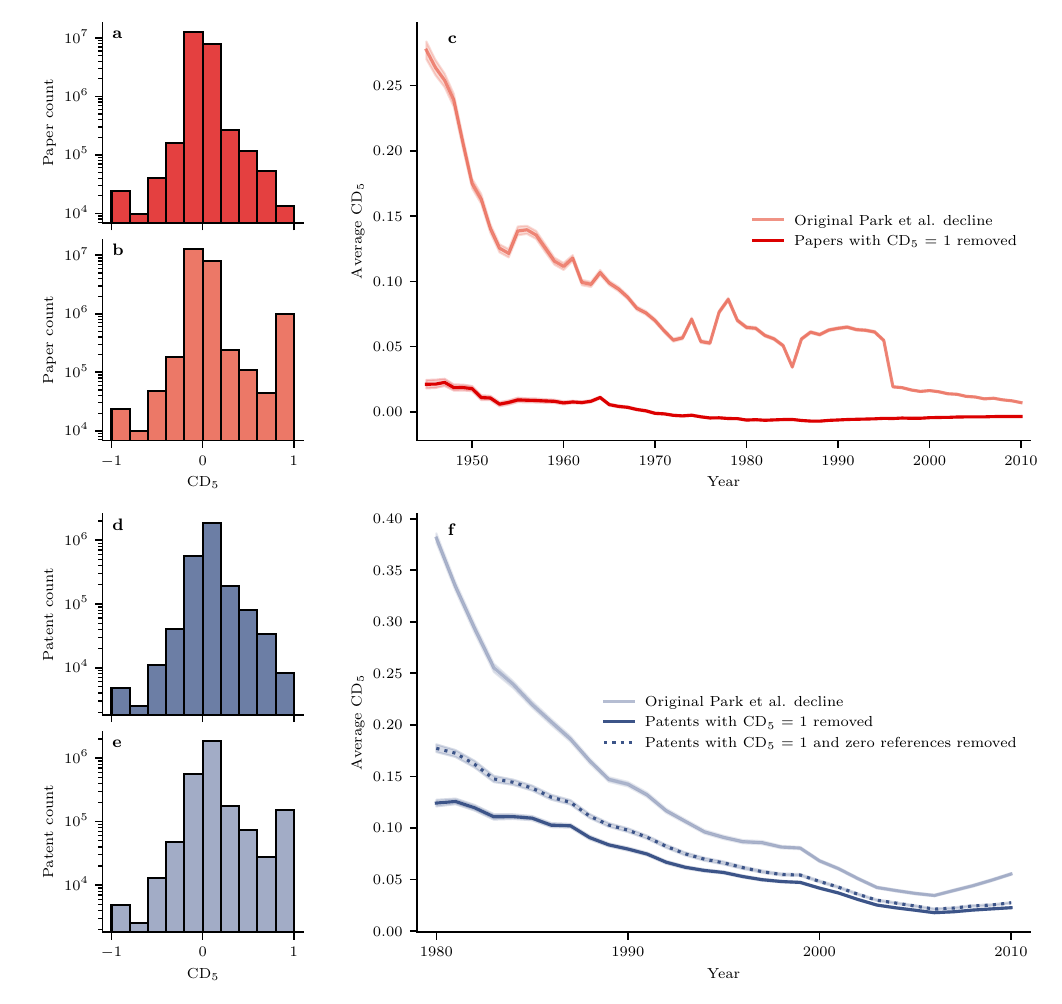}
\vspace{-0.5cm}
\caption{\textbf{$\vline$ Distribution of the $\mathbf{\mathrm{CD}_5}$ index with vs without the hidden outliers and its impact on the apparent decline of disruptive science and technology.} This figure shows that $\mathrm{CD}_5=1$ papers and patents are driving the reported decline in the disruptiveness of scientific and technological knowledge over time for the \textit{Web of Science} data source (with $22,479,429$ papers) and the \textit{PatentsView} data source (with $2,926,923$ patents). For \textit{PatentsView}, we also have access to sufficient metadata to exclude patents that make zero references, similarly impacting the decline. \textbf{a}, The distribution of the $\mathrm{CD}_5$ index for papers in \textit{Web of Science} as presented in Park et al.~\citep{park2023papers}, created using the binwidth parameter in \textit{seaborn 0.11.2}. This version of the library contains a bug regarding silently dropping the largest data points ($1$ in this case) when specifying the binwidth parameter \citep{seaborn2023pullrequest}. \textbf{b,} The correct histogram for papers when using the bins parameter in \textit{seaborn 0.11.2}. A peak at $\mathrm{CD}_5=1$ is revealed with $972,161$ additional papers. \textbf{c,} The time evolution of the average $\mathrm{CD}_5$ index for papers. When dropping the hidden outliers with $\mathrm{CD}_5=1$, the decline in disruptiveness almost completely disappears. The shaded bands correspond to $95\%$ confidence intervals. Finally, note that the curve without $\mathrm{CD}_5=1$ papers corresponds to (a), the histogram presented in Park et al.~\citep{park2023papers}. \textbf{d--f,} The equivalent plots for \textit{PatentsView} revealing $142,362$ additional patents with $\mathrm{CD}_5=1$. When dropping the outliers with $\mathrm{CD}_5=1$, the decline in disruptiveness reduces substantially. Unlike \textit{Web of Science}, the \textit{PatentsView} data source provided sufficient metadata to exclude patents with zero references, similarly impacting the data as removing outliers with $\mathrm{CD}_5=1$ (Fig. \ref{fig2} and Extended Data Fig. \ref{Extendedfig2}). Finally, note again that the curve without $\mathrm{CD}_5=1$ patents corresponds to (d), the histogram presented in Park et al.~\citep{park2023papers}.}
\label{fig1}
\end{figure}

Fig.~\ref{fig1}\textbf{a,d} reproduces the $\mathrm{CD}_5$ index distributions for papers and patents as presented in Park et al.~\citep{park2023papers} (Extended Data Fig.~1\textbf{a,c} in \citep{park2023papers}). A bug in the \textit{seaborn 0.11.2} plotting software~\citep{seaborn2023pullrequest}, used by Park et al.~\citep{park2023papers}, silently drops the largest data points in the histograms. Therefore, these histograms do not show the papers and patents with $\mathrm{CD}_5=1$. Using correct plot settings, Fig.~\ref{fig1}\textbf{b,e} reveals the additional $972,161$ papers and $142,362$ patents, with $\mathrm{CD}_5=1$. However, these hidden outliers were included in the main analysis in \citep{park2023papers}: the evaluation of the disruption versus time. Fig.~\ref{fig1}\textbf{c,f} shows that the decline in the disruptiveness of scientific (resp. technological) knowledge over time is negated (resp. substantially reduced) when these outliers are excluded.

The origin of these data points, and our reason for calling them outliers, can be found in their metadata. For patents, these are publicly available in the \textit{PatentsView} data source. As an open source alternative for \textit{Web of Science}, we use \textit{SciSciNet}~\citep{lin2023sciscinet}, an equivalent citation network with $39,888,199$ papers, for which we could replicate the above observation that $\mathrm{CD}_5=1$ papers are responsible for the temporal decline in disruption (Extended Data Fig.~\ref{Extendedfig1}). Extended Data Fig.~\ref{Extendedfig2}\textbf{a,d} shows the number of references made in each of those papers or patents according to the data source. For \textit{SciSciNet} and \textit{PatentsView}, we find that $97\%$ and $78\%$ of the $\mathrm{CD}_5=1$ papers and patents make zero references, respectively. Extended Data Fig.~\ref{Extendedfig2}\textbf{b,e} shows that within the $\mathrm{CD}_5 = 1$ category, the proportion of patents and papers that makes zero references is stable over time. Importantly, Extended Data Fig.~\ref{Extendedfig2}\textbf{c,f} reveals that the relative frequency of patents and papers with zero references and $\mathrm{CD}_5=1$ decreases over time, mirroring the decline in disruptiveness reported in \citep{park2023papers}. A second proof confirming the above mechanism is found by removing papers (Extended Data Fig.~\ref{Extendedfig1}\textbf{c}) and patents (Fig.~\ref{fig1}\textbf{f}) with $\mathrm{CD}_5=1$ and zero references, which has a similar effect on the reported decline compared to removing all hidden $\mathrm{CD}_5=1$ outliers, making the decline of disruptive science (technology) disappear (decrease).

One could argue that the consequences of the plotting mistake should have been caught by the robustness checks aimed at controlling for changing citation patterns over time~\citep{park2023papers}. We show below why both the regression adjustment and the Monte-Carlo simulations failed to do so.

\begin{figure}[ht!]%
\centering
\includegraphics[width=1.0\textwidth]{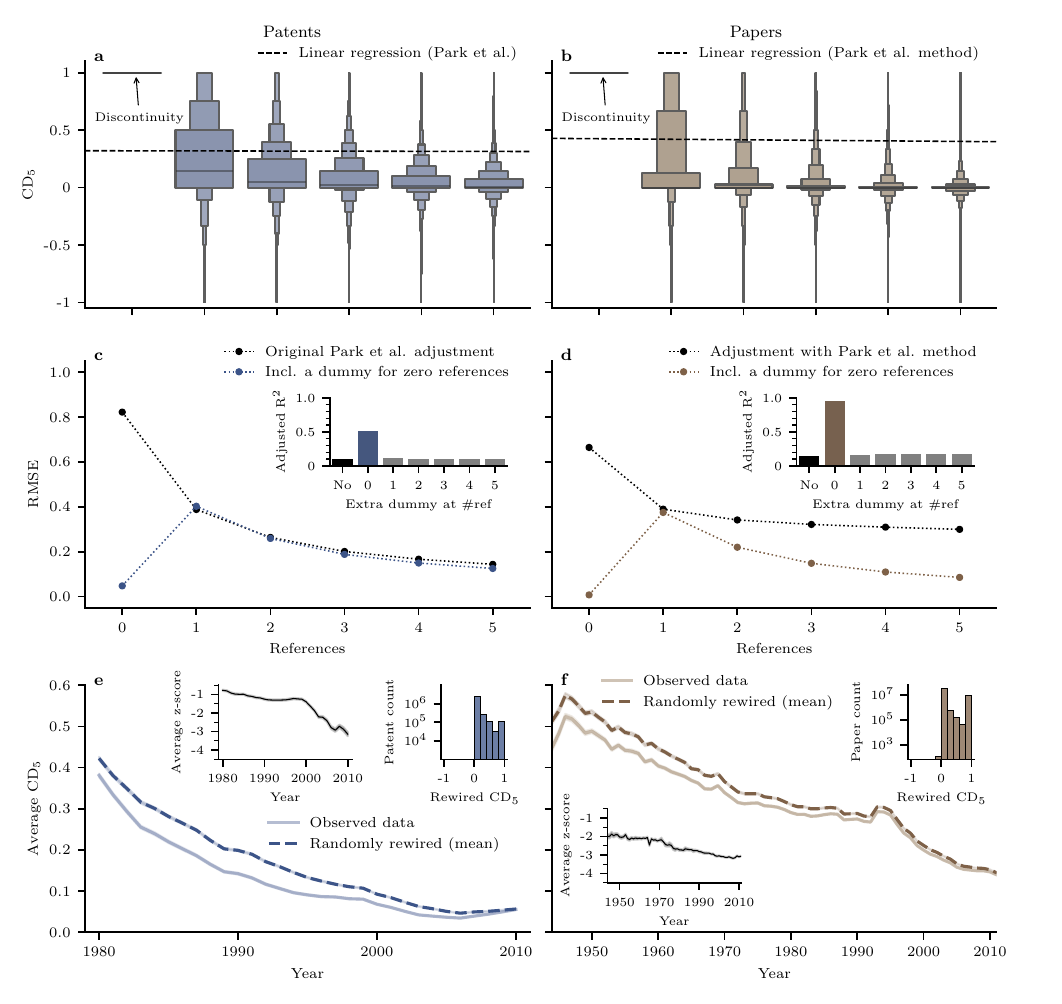}
\vspace{-0.65cm}
\caption{
\textbf{$\vline$ The reason why the robustness checks in Park et al.~\citep{park2023papers} failed to detect the consequences of the hidden outliers.}
This figure displays how the Park et al.~\citep{park2023papers} regression adjustment (models $4$ and $8$ in Supplementary Table $1$ in \citep{park2023papers}) fails to control for the discontinuous effect of zero references and that randomly rewired citation networks exhibit a similar temporal decline of $\mathrm{CD}_5$. Results are shown for \textit{PatentsView} (\textbf{a}, \textbf{c}, \textbf{e}; $n=2,926,923$ patents) using the original Park et al.~\citep{park2023papers} data and \textit{SciSciNet} \citep{lin2023sciscinet} (\textbf{b}, \textbf{d}, \textbf{f}; $n=39,888,199$ papers), replicating their \textit{Web of Science} analysis. Shaded bands correspond to $95 \%$ confidence intervals. 
\textbf{a,} The distribution of the $\mathrm{CD}_5$ per number of references is shown via letter-value plots which first identify the median, then extend boxes outward, each covering half of the remaining data~\citep{hofmann2017value}. Notably, in the case of zero references, the CD index is either one or remains undefined, causing a discontinuity. The marginal effect of references on $\mathrm{CD}_5$ shows that the regression adjustment of Park et al.~\citep{park2023papers} fails to account for this discontinuity.
\textbf{c,} The root mean squared errors (RMSE) show a pattern between the Park et al.~\citep{park2023papers} regression residuals and the number of references, showing that the model does not properly control for the discontinuous effect of zero references. Adding a dummy variable for zero references substantially improves the model fit as depicted by the adjusted $\mathrm{R}^2$, while a similar effect is not found for other reference dummy variables.
\textbf{e,} The average $\mathrm{CD}_5$ of the rewired patent networks (mean over ten runs) mirrors the decline of the observed network over time. This close similarity is the result of the one-to-one correspondence between zero reference patents within the observed and simulated networks, as evidenced by the peak at one in the histogram of the rewired $\mathrm{CD}_5$ shown in the inset plot. Finally, note that the gap between the observed $\mathrm{CD}_5$ values and those from the simulated networks is becoming smaller over time, which implies that the decline in the $z$ score found by Park et al.~\citep{park2023papers} and shown in the inset is the result of a decreasing standard deviation. \textbf{b}, \textbf{d}, \textbf{f}, The analogous, replicated plots for \textit{SciSciNet}.}
\label{fig2}
\end{figure}

First, Park et al.~\citep{park2023papers} proposed a linear regression to estimate an adjusted $\mathrm{CD}_5$ (models $4$ and $8$ in Supplementary Table 1 and Extended Data Fig.~8\textbf{b,e} in \citep{park2023papers}). The regression aims to predict the marginal effect of time by controlling for the number of references on the paper/patent level, together with fixed effects and additional control variables at the (sub-)field and year level. Notably, in the case of zero references to prior work, the CD index either equals the maximum value of one (when there is at least one citation) or remains undefined (when there are no citations). Fig.~\ref{fig2}\textbf{a,b} shows that the linear regression model~\citep{park2023papers} fails to control for this discontinuous effect of zero references. Fig.~\ref{fig2}\textbf{c,d} confirms that the regression errors (RMSE) peak at zero references. To control for the discontinuity, we extend the regression model from Park et al.~\citep{park2023papers} by including a dummy variable for zero references (Supplementary Table~\ref{tab:regression}). This results in an improved model fit, quantified by an adjusted explained variance ($\mathrm{R}^2$) increasing from $0.10$ to $0.52$ for patents and from $0.15$ to $0.95$ for papers. This improvement is not matched by the inclusion of a dummy variable for any other number of references (insets in Fig.~\ref{fig2}\textbf{c,d}). Supplementary Fig.~\ref{SupplementalFigure_Regression} shows that by explicitly controlling for the discontinuous effect of zero references, the decline in the disruptiveness of scientific (technological) knowledge is negated (reduced).

Second, Park et al.~\citep{park2023papers} conducted Monte Carlo simulations to control for the part of the observed decline caused by the general structure of the citation networks. They used a random rewiring algorithm \citep{uzzi2013atypical}, which preserves the publication years of the involved papers and patents and their number of forward citations and references, but randomly rearranges the citations between them. Based on an average $z$ score (Extended Data Fig. 8\textbf{c,f} in \citep{park2023papers}; see Supplementary Equation~\ref{definition_z_score} for more details), Park et al.~\citep{park2023papers} state that ``the observed $\mathrm{CD}_5$ values are lower than those from the simulated networks [\ldots] and the gap is widening.'' However, the temporal evolution of the CD index for the randomly rewired networks in Fig.~\ref{fig2}\textbf{e,f} instead shows a decline in disruption that almost perfectly mirrors the observed decline from the unaltered databases
(Supplementary Fig.~\ref{SupplementalFigure_Patents_rewiring}, \ref{SupplementalFigure_SciSciNet_rewiring}, \ref{SupplementalFigure_DBLPv14_rewiring}). Moreover, the gap between the observed $\mathrm{CD}_5$ values and those from the simulated networks narrows over time. The fact that the decline in disruption is present even in the randomly rewired networks can be explained by the degree-preserving nature of the rewiring algorithm, which induces a one-to-one correspondence between zero reference papers/patents in the original and rewired network, thus driving the decline in both networks. Consequently, random citation behaviour provides yet another proof that the relative decrease of zero reference patents and papers with $\mathrm{CD}_5=1$ per year (Extended Data Fig.~\ref{Extendedfig2}\textbf{c,f}) drives the reported decline in disruption.

Finally, we elucidate the source of this large quantity of papers and patents without references by inspecting $100$ randomly extracted patents and papers with zero references and $\mathrm{CD}_5=1$. We find that $98\%$ of the patent sample and $93\%$ of the paper sample do make references in their original PDF, indicating that most of the $\mathrm{CD}_5=1$ patents and papers with zero references should be treated as artefacts of the respective data sources rather than meaningful indicators of disruptive science and technology (Supplementary Tables~\ref{tab:summary_sample}-\ref{tab:patent_details}). While database errors in general do not only affect papers and patents with zero recorded references, they are especially problematic for these data entries, as having zero references causes a discontinuity in the CD index (Fig.~\ref{fig2}\textbf{a,b}). Therefore, it is best practice to exclude zero reference papers and patents prior to further analysis. Indeed, many recent Science of Science publications~\citep{wu2019large, lin2023sciscinet, lin2023remote} set the CD indices of papers that make zero references to non-defined.

We verified that our observations do not depend on the specific data source, the category within the respective data source, the choice of forward citation window, or the normalized CD indices (Extended Data Fig.~\ref{Extendedfig3}, Supplementary Figs.~\ref{SupplementalFig_DBLPv14}-\ref{Supplementalfig_disruption_indices}).

In summary, we revealed in three different ways that the decline in disruption, presented in Park et al.~\citep{park2023papers}, is driven by papers and patents with zero references and $\mathrm{CD}_5=1$. They remained hidden in the histograms, which the robustness checks failed to catch. Most of these papers and patents correspond to erroneous database entries. The curves showing how average CD indices have evolved, plotted in Park et al.~\citep{park2023papers}, therefore, do not track declining disruption of scientific and technological work, but rather trace how metadata quality has increased over time.

\clearpage
\backmatter




\bmhead{Acknowledgments}
We thank M. Park, E. Leahey, and R.J. Funk for making their code and source material public, and for responding to a previous version of this report, highlighting the different robustness checks in their manuscript.
We thank S.J. Klein (MIT) for helpful advice. We thank M. Waskom for maintaining the \textit{seaborn} library open source which allowed us to quickly identify the bug in the plotting of the histograms. \\\\
\textbf{Funding:} Work was supported by the Research Council (OZR) of the VUB.\\
\textbf{Competing interests:} The authors declare no competing interests.\\
\textbf{Ethics approval:} Not applicable.\\
\textbf{Consent to participate:} Not applicable.\\
\textbf{Consent for publication:} Not applicable.\\
\textbf{Availability of data and materials:} The \textit{Web of Science} and the \textit{PatentsView} data for the study was retrieved from the public repository (\url{https://doi.org/10.5281/zenodo.7258379}) made publicly available by Park et al.~\citep{park2023papers}. The \textit{SciSciNet} data source \citep{lin2023sciscinet} (\url{https://springernature.figshare.com/collections/SciSciNet_A_large-scale_open_data_lake_for_the_science_of_science_research/6076908/1}) and the \textit{DBLP-Citation-network V14} \citep{tang2008arnetminer} (\url{https://www.aminer.org/citation}) are publicly available to download. The data for the reanalysis we performed on these two datasets are publicly available at \url{https://github.com/VincentHolst/reanalysis_declining_disruption}. \\
\textbf{Code availability:} The re-analysis code for this publication is publicly available at \url{https://github.com/VincentHolst/reanalysis_declining_disruption}. It is based on the original analysis code (\url{https://doi.org/10.5281/zenodo.7258379}) made publicly available by Park et al.~\citep{park2023papers} and uses the same software packages as Park et al.~\citep{park2023papers}, that is \textit{pandas 1.4.3}, \textit{numpy 1.23.1}, \textit{matplotlib 3.5.2} and \textit{seaborn 0.11.2}. To replicate the regression table, we used StataMP v.18.0 (reghdfe v.6.12.3).\\
\textbf{Authors' contributions:} V.H. and V.G. were responsible for the main idea of the study. V.H. discovered the mistake in the original analysis (outliers hidden). A.A. discovered the reason why the mistake was made (seaborn update). V.H., A.A., and V.G. designed the analysis. A.A. replicated the regression analysis. V.H. and F.T. implemented the random rewiring algorithm. V.H. and F.T. tested the robustness of the results with additional data and analyses. V.H. and F.T. designed the final figures. S.W. independently reviewed the results. All authors discussed the results and collaboratively drafted and revised the manuscript.



\clearpage
\begin{appendices}

\tocless\section{Extended Data}
\begin{figure}[ht!]%
\centering
\includegraphics[width=1.0\textwidth]{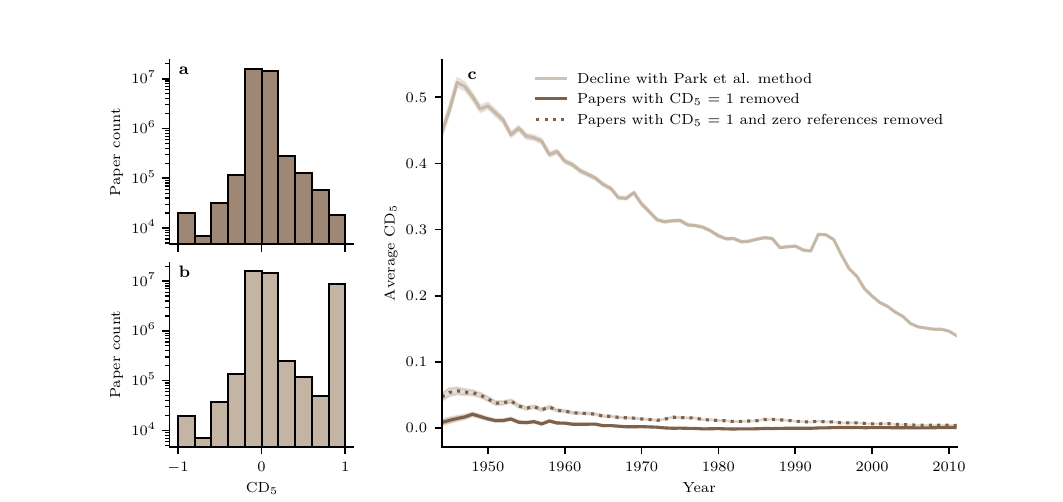}
\vspace{-0.5 cm}
\caption{\textbf{$\vline$ Distribution of the $\mathrm{CD}_5$ index with vs without the hidden outliers and its impact on the disruptiveness for the \textit{SciSciNet} data source.} 
This figure replicates the observation that papers with $\mathrm{CD}_5 = 1$ are driving the decline in disruptive science for the \textit{SciSciNet} data source \citep{lin2023sciscinet} (with $39,888,199$ papers between $1944$ and $2011$), which originated from the \textit{Microsoft Academic Graph}. \textbf{a}, The distribution of the $\mathrm{CD}_5$ index for \textit{SciSciNet}, created using the binwidth parameter in \textit{seaborn 0.11.2}. Here again, the largest data points are hidden. \textbf{b,} The correct histogram of the underlying dataset. A peak at $\mathrm{CD}_5=1$ is revealed, corresponding to $8,861,343$ additional papers. \textbf{c,} The time evolution of the average $\mathrm{CD}_5$ index. When dropping the outliers with $\mathrm{CD}_5=1$, the decline in disruptiveness is negated. Excluding papers with zero references impacts the data similarly (Fig.~\ref{fig2} and Extended Data Fig.~\ref{Extendedfig2}). The shaded bands correspond to $95\%$ confidence intervals. Moreover, the curve with papers with $\mathrm{CD}_5 = 1$ omitted is the curve corresponding to the histogram~(a).}\label{Extendedfig1}
\end{figure}

\begin{figure}[ht]%
\centering
\includegraphics[width = 1.0\textwidth]{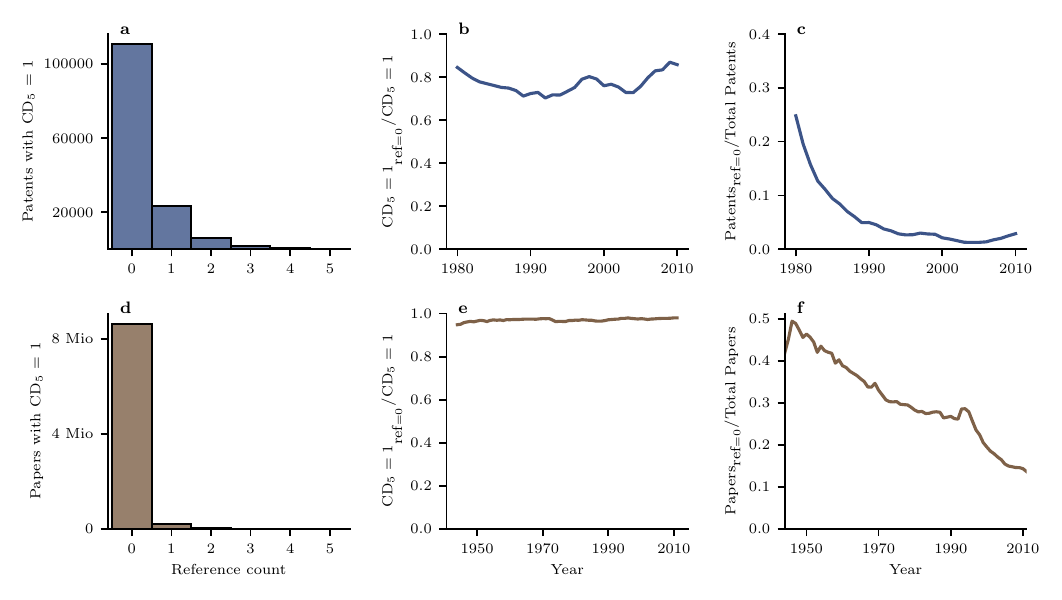}
\vspace{-0.5 cm}
\caption{\textbf{$\vline$ Papers and patents with $\mathrm{CD}_5=1$ predominantly make zero references.} This figure displays that most papers in the \textit{SciSciNet} data source \citep{lin2023sciscinet} ($n=39,888,199$) and most patents in the \textit{PatentsView} data source ($n=2,926,923$) with $\mathrm{CD}_5=1$ have zero references. \textbf{a,} Our analysis shows that \textit{PatentsView} contains $142,362$ patents with $\mathrm{CD}_5=1$ between $1980$ and $2010$, of which $78 \: \%$ appear in the database with zero references. \textbf{b,} Within the category of patents with $\mathrm{CD}_5 = 1$, the relative frequency of patents with zero references is stable between $1980$ and $2010$. \textbf{c,} The relative frequency of patents with $\mathrm{CD}_5$ index exactly equal to one and zero references is decreasing over time. Therefore, a substantial part of the reported decline in the disruptiveness of technological knowledge over time can be attributed to a relatively increasing metadata quality over time. It is also intriguing to note how well the shape of this curve resembles the shape of the top curve shown in Fig.~\ref{fig1}\textbf{f}. \textbf{d,} \textit{SciSciNet}~\citep{lin2023sciscinet} shows a similar behaviour with $8,861,343$ papers having $\mathrm{CD}_5=1$ between $1944$ and $2011$, of which $97 \: \%$ appear in the database with zero references. \textbf{e,} Within the category of papers with $\mathrm{CD}_5=1$, the relative frequency of papers with zero references is stable between $1944$ and $2011$. \textbf{f,} The relative frequency of papers with $\mathrm{CD}_5$ index exactly equal to one and zero references is decreasing over time. Therefore, a substantial part of the observed decline in the disruptiveness of scientific knowledge over time can be attributed to a relatively increasing metadata quality over time. It is also intriguing to note how well the shape of this curve resembles the shape of the top curve shown in Extended Data Fig.~\ref{Extendedfig1}\textbf{c}.
}\label{Extendedfig2}
\end{figure}

\begin{figure}[ht!]%
\centering
\includegraphics[width=1.0\textwidth]{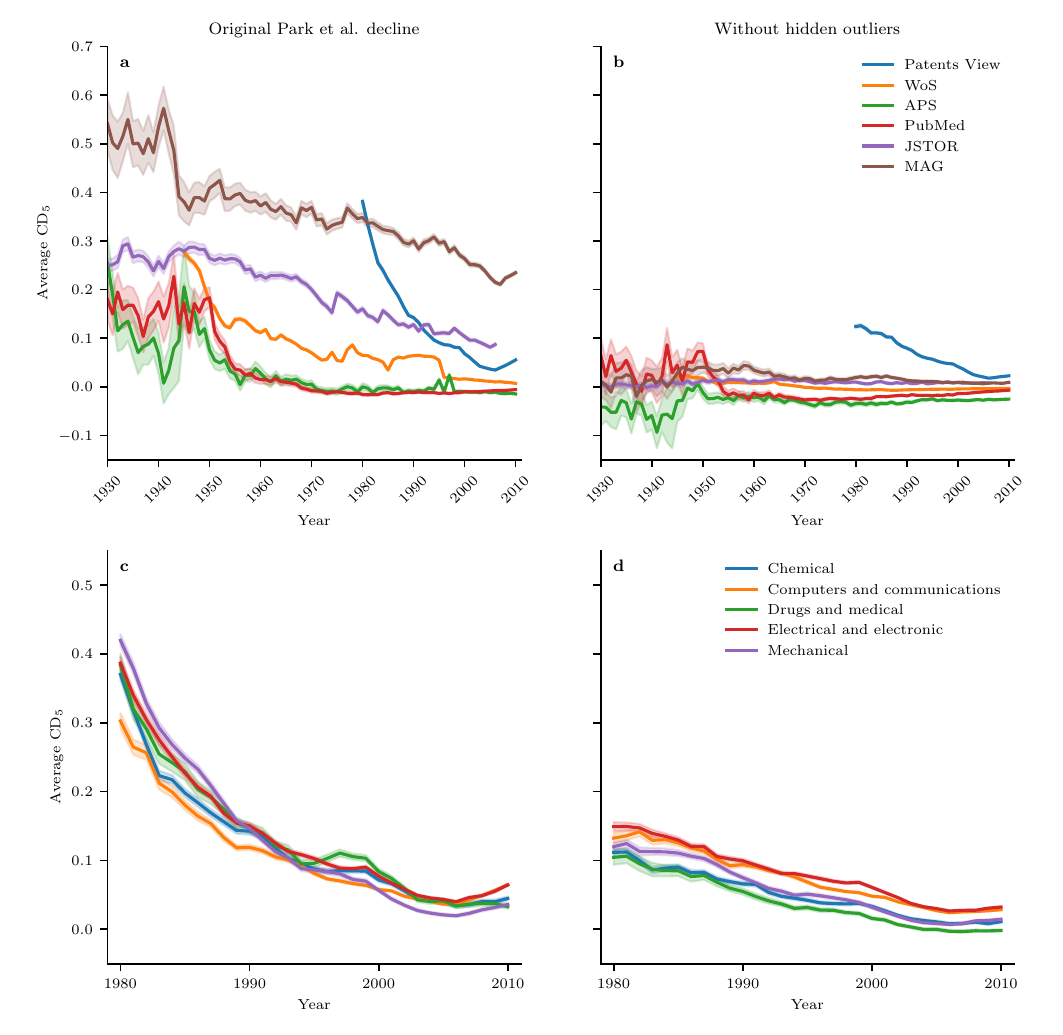}
\vspace{-0.5 cm}
\caption{\textbf{$\vline$ Across various data sources and within different categories, papers and patents with $\mathrm{CD}_5=1$ are driving the decline in the disruptiveness in scientific and technological knowledge over time.} This figure displays the average $\mathrm{CD}_5$ index over time for six data sources and five different patent categories. The data sources are \textit{JSTOR} ($1,588,088$ papers), the \textit{American Physical Society} corpus ($461,359$ papers), \textit{Microsoft Academic Graph} (random sample of $1,000,000$ papers), and \textit{PubMed} ($1,563,211$ papers). For reference, the \textit{Web of Science} ($22,479,429$ papers) and \textit{PatentsView} ($2,926,923$ patents) data sources are also included. The patent categories are \textit{Chemical} ($517,964$ patents), \textit{Computers and communications} ($748,849$ patents), \textit{Drugs and medical} ($321,449$ patents), \textit{Electrical and electronic} ($734,769$ patents) and \textit{Mechanical} ($603,892$ patents). Shaded bands correspond to $95\%$ confidence intervals. \textbf{a}, The temporal evolution of the average $\mathrm{CD}_5$ index for different data sources as presented in Park et al.~\citep{park2023papers} (Extended Data Fig.~6 in \citep{park2023papers}). \textbf{b}, The time evolution of the average $\mathrm{CD}_5$ index for different data sources after removing the outliers with $\mathrm{CD}_5 = 1$ from the data sources. For all mentioned data sources that encompass papers, the decline in the disruptiveness almost completely disappears. For the \textit{PatentsView} data source, the decline in the disruptiveness also reduces notably. \textbf{c}, The time evolution of the average $\mathrm{CD}_5$ index for different patent categories as presented in Park et al.~\citep{park2023papers} (Fig.~2\textbf{b} in \citep{park2023papers}). \textbf{d}, The time evolution of the average $\mathrm{CD}_5$ index for different patent categories after removing the outliers with $\mathrm{CD}_5 = 1$ from the categories. We see that the decline in disruptiveness reduces similarly across all five categories.}\label{Extendedfig3}
\end{figure}

\end{appendices}

\clearpage
\part*{SI Guide for ``Dataset Artefacts are the Hidden Drivers of the Declining Disruptiveness in Science''}
Vincent Holst, Data Analytics Laboratory, Vrije Universiteit Brussel \\
Andres Algaba, Data Analytics Laboratory, Vrije Universiteit Brussel \\ 
Floriano Tori, Data Analytics Laboratory, Vrije Universiteit Brussel \\
Sylvia Wenmackers, Centre for Logic and Philosophy of Science (CLPS), KU Leuven \\
Vincent Ginis, Data Analytics Laboratory, Vrije Universiteit Brussel \& School of 

Engineering and Applied Sciences, Harvard University 

\section*{Supplementary Methods}
Here, we provide additional mathematical justifications, supplementary tables and supplementary figures for our reanalysis of Park et al.~\citep{park2023papers}.
\tableofcontents
\newpage
\part*{Supplementary Information}\label{sec:supplemental}
\setcounter{figure}{0}
\renewcommand{\thefigure}{S\arabic{figure}}

\setcounter{table}{0}
\renewcommand{\thetable}{S\arabic{table}}

\setcounter{equation}{0}
\renewcommand{\theequation}{S\arabic{equation}}

\setcounter{section}{0}
\renewcommand{\thesection}{S\arabic{section}.}

\section{~The CD index}
Let $\mathcal{G}=(V, E)$ be a directed citation network. The set $V$ corresponds to the papers in $A$ and $E$ corresponds to the citations between the papers. The adjacency matrix $A=(a_{ij})$ of $\mathcal{G}=(V, E)$ is given by $a_{ij}=1$ if and only if paper $i$ cites paper $j$, and $a_{ij}=0$ otherwise. Every paper $i \in V$ is assigned a publication date $d_i$, usually given in datetime format. The directed citation network $\mathcal{G}$ possesses a temporal structure: if paper $i$ cites paper $j$, we have $d_i > d_j$, i.e. paper $i$ was published after paper $j$. 

Let $i \in V$ be a focal paper with publication date $d_i$ and $t \in \mathbb{N}^{+}=\{1, 2, 3, ...\}$ a forward citation window. Let $U \subset V$ be the papers published between $(d_i, d_i+ t \text{ years}]$, e.g., if $d_i$ is equal to 1984-01-01, then $U$ encompasses all papers published after $d_i$ until 1989-01-01. We consider the following sets:
\begin{align} \label{definition_cd_1}
    & F := \{ j \in U \vert \: j \text{ cites the \underline{focal} paper } i \text{ but none of its references} \}, \notag \\  
    & B := \{ j \in U \vert \: j \text{ cites \underline{both} the focal paper } i \text{ and at least one of its references} \}, \\ 
     & R := \{ j \in U \vert \: j \text{ does not cite the focal paper } i \text{ but at least one of its \underline{references}} \}. \notag
\end{align}
Let $N_F= \vert F \vert, \: N_B = \vert B \vert$ and $N_R= \vert R \vert$. Then
the $\mathrm{CD}_t$ index \citep{funk2017dynamic} of paper $i$ is given by:
\begin{equation} \label{definition_cd_2}
    \mathrm{CD}_t=\frac{N_F-N_B}{N_F+N_B+N_R}.
\end{equation}
If paper $i$ has zero references to prior work, the sets $B$ and $R$ are empty by default and it is easy to see that $\mathrm{CD}_t=(N_F-N_B)/(N_F+N_B+N_R)=N_F/N_F$ is either exactly equal to one (if $F$ is not empty, i.e. if $i$ receives at least one forward citation within $t$ years after publication) or remains undefined (if $F$ is empty, i.e. if $i$ receives no forward citation within $t$ years after publication).

The \textit{DBLP-Citation-network V14} \citep{tang2008arnetminer} provides the publication date only in YYYY format. If the focal paper $i$ is published in a given year $d_i$, we considered the subsets $U \subset V$ of papers published between $[d_i+1, d_i+ 5]$ and $W \subset V$ of papers published between $[d_i, d_i+ 5]$ as follows in the calculation of the $\mathrm{CD}_5$ index:
\begin{align} \label{DBLP_cd5}
    & F := \{ j \in W \vert \: j \text{ cites the \underline{focal} paper } i \text{ but none of its references} \}, \notag \\  
    & B := \{ j \in W \vert \: j \text{ cites \underline{both} the focal paper } i \text{ and at least one of its references} \}, \\ 
     & R := \{ j \in U \vert \: j \text{ does not cite the focal paper } i \text{ but at least one of its \underline{references}} \}. \notag
\end{align}

\clearpage
\section{~Regression adjustment}
\begin{figure}[ht!]%
\centering
\includegraphics[width=1.0\textwidth]{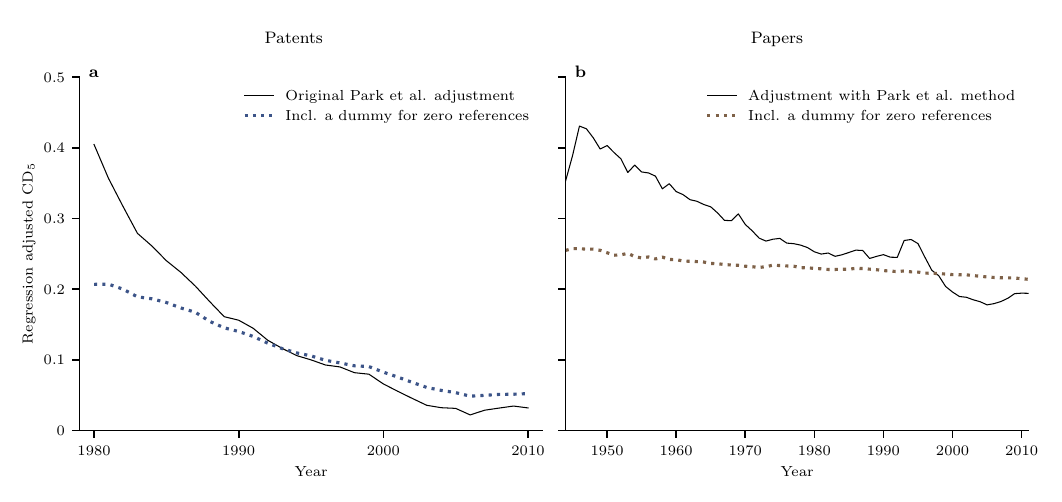}
\vspace{-0.5cm}
\caption{
\textbf{$\vline$ Inclusion of the zero references dummy variable to properly control for the corresponding discontinuity in the $\mathrm{CD}_5$ largely negates the decline of disruptive science.}
This figure displays that the inclusion of the zero references dummy variable to account for the corresponding discontinuity in the $\mathrm{CD}_5$ (Fig.~\ref{fig2}\textbf{a,b}) in the regression adjustment substantially reduces the decline for the \textit{PatentsView} data source (with $2,926,923$ patents, models ($3$) and ($4$) in Supplementary Table \ref{tab:regression}) and largely negates the decline for the \textit{SciSciNet} data source~\citep{lin2023sciscinet} (with $39,888,199$ papers, models ($1$) and ($2$) in Supplementary Table \ref{tab:regression}). \textbf{a,} As displayed in Fig.~\ref{fig2}\textbf{a,c}, the linear regression conducted by Park et al.~\citep{park2023papers} for \textit{PatentsView} (model $8$ from Supplementary Table 1 in \citep{park2023papers}) fails to control for the discontinuous effect of zero references. This is shown by including a dummy variable explicitly controlling for zero references, resulting in a substantial increase in the adjusted $\mathrm{R}^2$, an effect not observed for any other number of references. Here, we show that explicitly controlling for the discontinuous effect of zero references substantially reduces the temporal decline of the regression adjusted $\mathrm{CD}_5$. \textbf{b,} The analogous analysis for \textit{SciSciNet}. Here, we replicate the Park et al.~\citep{park2023papers} regression model conducted for \textit{Web of Science} (model $4$ from Supplementary Table $1$ in \citep{park2023papers}). Notably, explicitly controlling for the discontinuous effect of zero references largely negates the temporal decline of the regression adjusted $\mathrm{CD}_5$.
}
\label{SupplementalFigure_Regression}
\end{figure}

Park et al.~\citep{park2023papers} use a linear regression to control for potential changes in citation patterns by including the number of references on the paper/patent level, together with fixed effects and additional control variables at the (sub-)field and year level. We also include a zero references dummy variable, which is equal to one if the paper/patent has zero references and zero else, to explicitly control for the discontinuous effect of zero references (Fig.~\ref{fig2}\textbf{a,b}). The regression looks as follows:
\begin{align} \label{Regression equation}
& \mathrm{CD}_{5_{i,t(i),k(i)}} = \; \alpha + \underbrace{\sum_{t=1}^{T-1} \theta_{t} \; \text{year}_{t(i)}}_{\substack{\text{time} \\ \text{fixed effects}}} + \underbrace{\sum_{k=1}^{K-1} \delta_k \; \text{(sub-)field}_{k(i)}}_{\substack{\text{(sub-)field} \\ \text{fixed effects}}} + \underbrace{\beta_1 \; \text{\#references}_{i}}_{\substack{\text{paper/patent} \\ \text{level control}}} \nonumber \\
& + \; \underbrace{\gamma_1 \; \text{\#patents/papers}_{t(i), k(i)} + \gamma_2 \; \overline{\text{\#references}}_{t(i), k(i)} + \gamma_3 \; \overline{\text{\#authors/inventors}}_{t(i), k(i)}}_{\substack{\text{(sub-)field and year} \\ \text{level controls}}} \nonumber \\
& + \; \underbrace{\zeta \; \text{zero references}_i}_{\substack{\text{zero references} \\ \text{dummy variable}}} + \; \varepsilon_{i},
\end{align}
where $i$, $t$, and $k$ denote the paper/patent, the publication/grant year, and (sub-)field, respectively. Moreover, $t(i)$ and $k(i)$ indicate that the publication/grant year $t$ and (sub-)field $k$ depend on the paper/patent $i$, $\#$ denotes ``the number of,'' and $\overline{x}$ is the average of a variable $x$. The time and (sub-)field fixed effects can be estimated by including dummy variables for the publication/grant years and (sub-)fields, which are equal to one if the year or field is equal to the year or field of the current paper/patent and zero else, with the exclusion of a reference category (i.e., $T-1$ and $K-1$). Note that we include the zero references dummy variable to explicitly control for the discontinuous effect and to show the consequences for the main findings of Park et al.~\citep{park2023papers} by displaying the regression adjusted $\mathrm{CD}_5$ (Supplementary Fig.~\ref{SupplementalFigure_Regression} and Supplementary Table~\ref{tab:regression}). Since we only want to show that the regression model of Park et al.~\citep{park2023papers} does not control for the discontinuity of zero references, we do not make any further changes to the regression model, despite other potential improvements, such as taking the natural logarithm of the number of references as a control variable~\citep{ruan2021rethinking}.

\newpage

\newpage
\begin{table}[ht]%
\centering
\caption{Regression models for adjusting $\mathrm{CD}_5$.}
\label{tab:regression}
\footnotesize 
\renewcommand{\arraystretch}{0.01} 
\setlength{\tabcolsep}{3.5pt} 
\begin{tabular}{{@{\extracolsep{4pt}}l C{2cm} C{2cm} C{2cm} C{2cm}@{}}}
& \multicolumn{2}{c}{SciSciNet} & \multicolumn{2}{c}{PatentsView} \\
\cmidrule{2-3} \cmidrule{4-5} 
Variables & (1) & (2) & (3) & (4) \\
\midrule
Year=1945 & $0.04^{***}$ & $0.00$ & \\
Year=1946 & $0.08^{***}$ & $0.00$ & \\
Year=1947 & $0.07^{***}$ & $0.00$ & \\
Year=1948 & $0.06^{***}$ & $0.00$ & \\
Year=1949 & $0.05^{***}$ & $0.00$ & \\
Year=1950 & $0.05^{***}$ & $-0.00^{*}$ & \\
Year=1951 & $0.04^{***}$ & $-0.01^{***}$ & \\
Year=1952 & $0.03^{***}$ & $-0.01^{***}$ & \\
Year=1953 & $0.01^{***}$ & $-0.00^{**}$ & \\
Year=1954 & $0.02^{***}$ & $-0.01^{***}$ & \\
Year=1955 & $0.01^{***}$ & $-0.01^{***}$ & \\
Year=1956 & $0.01^{**}$ & $-0.01^{***}$ & \\
Year=1957 & $0.01$ & $-0.01^{***}$ & \\
Year=1958 & $-0.01^{***}$ & $-0.01^{***}$ & \\
Year=1959 & $-0.00$ & $-0.01^{***}$ & \\
Year=1960 & $-0.02^{***}$ & $-0.01^{***}$ & \\
Year=1961 & $-0.02^{***}$ & $-0.02^{***}$ & \\
Year=1962 & $-0.03^{***}$ & $-0.02^{***}$ &\\
Year=1963 & $-0.03^{***}$ & $-0.02^{***}$ &\\
Year=1964 & $-0.03^{***}$ & $-0.02^{***}$ &\\
Year=1965 & $-0.04^{***}$ & $-0.02^{***}$ &\\
Year=1966 & $-0.05^{***}$ & $-0.02^{***}$ &\\
Year=1967 & $-0.06^{***}$ & $-0.02^{***}$ &\\
Year=1968 & $-0.06^{***}$ & $-0.02^{***}$ &\\
Year=1969 & $-0.05^{***}$ & $-0.02^{***}$ &\\
Year=1970 & $-0.06^{***}$ & $-0.02^{***}$ &\\
Year=1971 & $-0.07^{***}$ & $-0.02^{***}$ &\\
Year=1972 & $-0.08^{***}$ & $-0.02^{***}$ &\\
Year=1973 & $-0.09^{***}$ & $-0.02^{***}$ &\\
Year=1974 & $-0.08^{***}$ & $-0.02^{***}$ &\\
Year=1975 & $-0.08^{***}$ & $-0.02^{***}$ &\\
Year=1976 & $-0.09^{***}$ & $-0.02^{***}$ &\\
Year=1977 & $-0.09^{***}$ & $-0.02^{***}$ &\\
Year=1978 & $-0.09^{***}$ & $-0.02^{***}$ &\\
Year=1979 & $-0.09^{***}$ & $-0.03^{***}$ &\\
Year=1980 & $-0.10^{***}$ & $-0.03^{***}$ &\\
Year=1981 & $-0.10^{***}$ & $-0.03^{***}$ & $-0.05^{***}$ & $0.00$ \\
Year=1982 & $-0.10^{***}$ & $-0.03^{***}$ & $-0.09^{***}$ & $-0.01^{***}$ \\
Year=1983 & $-0.11^{***}$ & $-0.03^{***}$ & $-0.13^{***}$ & $-0.02^{***}$ \\
Year=1984 & $-0.11^{***}$ & $-0.03^{***}$ & $-0.14^{***}$ & $-0.02^{***}$ \\
Year=1985 & $-0.10^{***}$ & $-0.03^{***}$ & $-0.16^{***}$ & $-0.03^{***}$ \\
Year=1986 & $-0.10^{***}$ & $-0.03^{***}$ & $-0.18^{***}$ & $-0.03^{***}$ \\
Year=1987 & $-0.10^{***}$ & $-0.03^{***}$ & $-0.20^{***}$ & $-0.04^{***}$ \\
Year=1988 & $-0.11^{***}$ & $-0.03^{***}$ & $-0.22^{***}$ & $-0.05^{***}$ \\
Year=1989 & $-0.11^{***}$ & $-0.03^{***}$ & $-0.24^{***}$ & $-0.06^{***}$ \\
Year=1990 & $-0.10^{***}$ & $-0.03^{***}$ & $-0.25^{***}$ & $-0.07^{***}$ \\
Year=1991 & $-0.11^{***}$ & $-0.03^{***}$ & $-0.26^{***}$ & $-0.07^{***}$ \\
Year=1992 & $-0.11^{***}$ & $-0.03^{***}$ & $-0.28^{***}$ & $-0.08^{***}$ \\
Year=1993 & $-0.09^{***}$ & $-0.03^{***}$ & $-0.29^{***}$ & $-0.09^{***}$ \\
Year=1994 & $-0.08^{***}$ & $-0.03^{***}$ & $-0.30^{***}$ & $-0.10^{***}$ \\
Year=1995 & $-0.09^{***}$ & $-0.03^{***}$ & $-0.30^{***}$ & $-0.10^{***}$ \\
Year=1996 & $-0.11^{***}$ & $-0.03^{***}$ & $-0.31^{***}$ & $-0.11^{***}$ \\
Year=1997 & $-0.13^{***}$ & $-0.03^{***}$ & $-0.32^{***}$ & $-0.11^{***}$ \\
Year=1998 & $-0.13^{***}$ & $-0.03^{***}$ & $-0.32^{***}$ & $-0.12^{***}$ \\
Year=1999 & $-0.15^{***}$ & $-0.03^{***}$ & $-0.33^{***}$ & $-0.12^{***}$ \\
Year=2000 & $-0.16^{***}$ & $-0.03^{***}$ & $-0.34^{***}$ & $-0.12^{***}$ \\
Year=2001 & $-0.16^{***}$ & $-0.04^{***}$ & $-0.35^{***}$ & $-0.13^{***}$ \\
Year=2002 & $-0.17^{***}$ & $-0.04^{***}$ & $-0.36^{***}$ & $-0.14^{***}$ \\
Year=2003 & $-0.17^{***}$ & $-0.04^{***}$ & $-0.37^{***}$ & $-0.15^{***}$ \\
Year=2004 & $-0.17^{***}$ & $-0.04^{***}$ & $-0.37^{***}$ & $-0.15^{***}$ \\
Year=2005 & $-0.18^{***}$ & $-0.04^{***}$ & $-0.37^{***}$ & $-0.15^{***}$ \\
Year=2006 & $-0.17^{***}$ & $-0.04^{***}$ & $-0.38^{***}$ & $-0.16^{***}$ \\
Year=2007 & $-0.17^{***}$ & $-0.04^{***}$ & $-0.38^{***}$ & $-0.16^{***}$ \\
Year=2008 & $-0.17^{***}$ & $-0.04^{***}$ & $-0.37^{***}$ & $-0.16^{***}$ \\
Year=2009 & $-0.16^{***}$ & $-0.04^{***}$ & $-0.37^{***}$ & $-0.16^{***}$ \\
Year=2010 & $-0.16^{***}$ & $-0.04^{***}$ & $-0.37^{***}$ & $-0.16^{***}$ \\
Year=2011 & $-0.16^{***}$ & $-0.04^{***}$ \\\\
$\beta_1$& $-4.73\mathrm{e}{-3}^{***}$ & $-2.73\mathrm{e}{-4}^{***}$ & $-1.11\mathrm{e}{-3}^{***}$ & $-6.46\mathrm{e}{-4}^{***}$ \\
$\gamma_1$ & $-3.20\mathrm{e}{-7}^{***}$ & $1.23\mathrm{e}{-8}^{***}$ & $3.17\mathrm{e}{-6}^{***}$ & $5.67\mathrm{e}{-7}^{***}$ \\
$\gamma_2$ & $5.94\mathrm{e}{-4}^{***}$ & $1.03\mathrm{e}{-4}^{***}$ & $9.61\mathrm{e}{-4}^{***}$ & $7.78\mathrm{e}{-4}^{***}$ \\
$\gamma_3$ & $1.23\mathrm{e}{-2}^{***}$ & $3.93\mathrm{e}{-3}^{***}$ & $2.64\mathrm{e}{-2}^{***}$ & $1.77\mathrm{e}{-2}^{***}$ \\\\
$\zeta$ & & $0.98^{***}$ & & $0.90^{***}$ \\\\
$\alpha$ & $0.43^{***}$ & $0.03^{***}$ & $0.32^{***}$ & $0.13^{***}$ \\
\midrule
(Sub-)field fixed effects  & yes & yes & yes & yes \\
\midrule
N & $39,888,199$ & $39,888,199$ & $2,926,923$ & $2,926,923$ \\
Adjusted $R^2$ & $0.15$ & $0.95$ & $0.10$ & $0.52$ \\
\bottomrule
\end{tabular}
\small
{\footnotesize Note: Model 3 is Model 8 from~\citep{park2023papers} (Suppl. Table 1). Model 1 replicates Model 4 from~\citep{park2023papers} (Suppl. Table 1) on SciSciNet instead of WoS. Models 2 and 4 control for zero references by including a dummy variable. Estimates are from an OLS-regression (Eq. \ref{Regression equation}) and significance levels are for a two-sided t-test with a $\text{H}_0$ of the regression coefficient being equal to zero (\textsuperscript{***}$p<0.01$, \textsuperscript{**}$p<0.05$, \textsuperscript{*}$p<0.1$}).
\end{table}

\clearpage
\newpage
\section{~Monte Carlo simulations} 
In their original manuscript, Park et al.~\citep{park2023papers} conducted Monte Carlo simulations to check if the observed decline of $\mathrm{CD}_5$ is caused by changes in the citation networks' general topology instead of societal processes. Therefore, Park et al.~\citep{park2023papers} used a random rewiring algorithm \citep{uzzi2013atypical} that preserves the topological structure, i.e. the in- and outdegree (resp. number of forward citations and references) of the involved papers and patents, and the age structure, i.e. the publications years of the involved papers and patents, but randomly rewires the citations between the involved papers and patents. 

Park et al.~\citep{park2023papers} described the rewiring algorithm as follows: if paper A cites paper B and paper C cites paper D, then the switch to paper A cites paper D and paper C cites paper B is retained if and only if (1) paper A and paper C (resp. paper B and paper D) have the same number of forward citations (resp. references) after the switch and (2) paper A and paper C (resp. paper B and paper D) were published in the same year. The random rewiring is repeated until $100 \cdot \# \textrm{edges}$ switches are retained.

\begin{figure}[ht!]%
\centering
\includegraphics[width=1.0\textwidth]{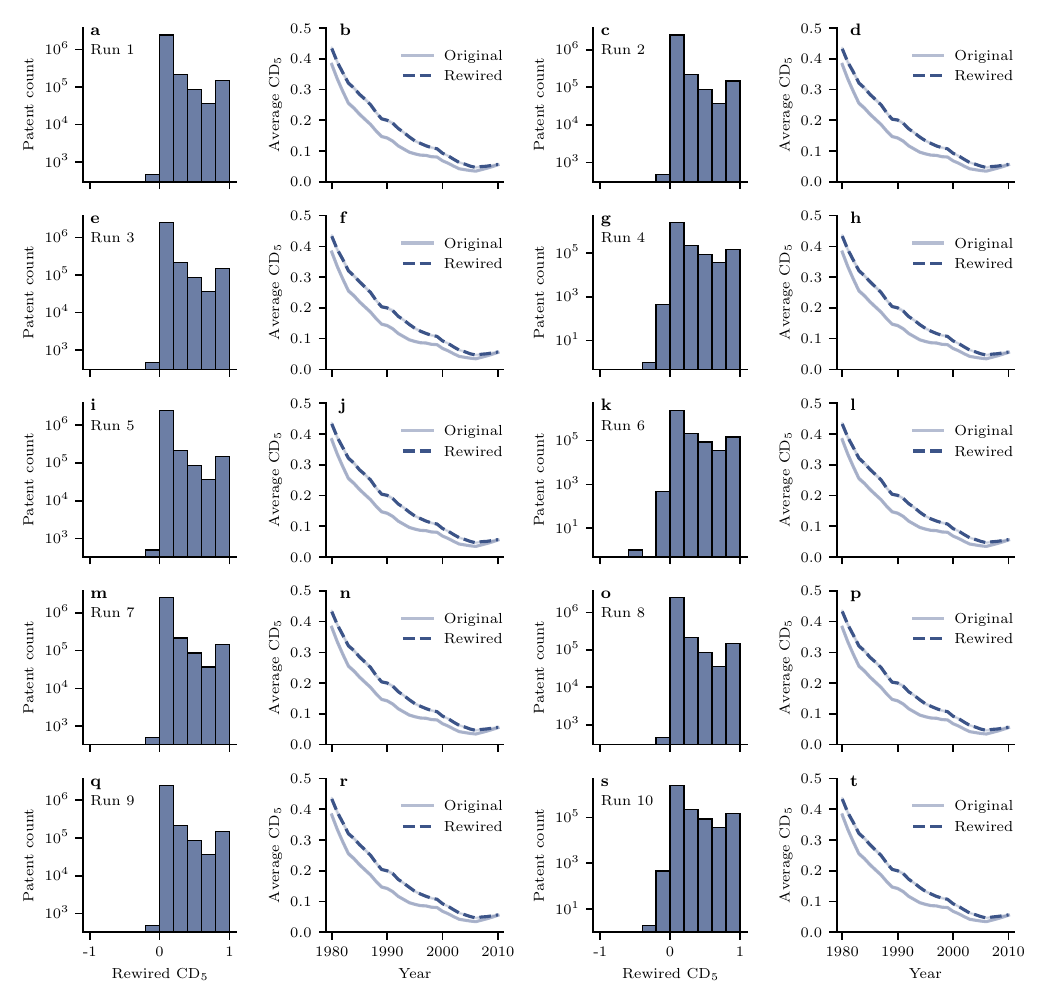}
\vspace{-0.5 cm}
\caption{\textbf{$\vline$ The temporal decline of the $\mathrm{CD}_5$ index for patents is mirrored by random citation behaviour supporting that the hidden outliers are driving the decline.} This figure displays the distribution (panels \textbf{a}, \textbf{c}, \textbf{e}, \textbf{g}, \textbf{i}, \textbf{k}, \textbf{m}, \textbf{o}, \textbf{q}, \textbf{s}) and time average (panels \textbf{b}, \textbf{d}, \textbf{f}, \textbf{h}, \textbf{j}, \textbf{l}, \textbf{n}, \textbf{p}, \textbf{r}, \textbf{t}) of the $\mathrm{CD}_5$ index for the ten randomly rewired \textit{PatentsView} data sources (with $2,926,923$ patents) from \citep{park2023papers}. Here, the random rewiring algorithm~\citep{uzzi2013atypical} used by Park et al.~\citep{park2023papers} preserves the in- and outdegree (resp. forward citations and references) and publication year of the involved patents. In particular, this induces a one-to-one correspondence between the zero reference patents in the rewired and original network. The shaded bands in the plots correspond to $95 \: \%$ confidence intervals. \textbf{a}, The distribution of the rewired $\mathrm{CD}_5$ for the first of the ten randomly rewired patent networks shows that the algorithm used by Park et al.~\citep{park2023papers} boosts CD index values. This is unsurprising, as the CD index measures triadic closure and randomly rewiring a sparse citation network naturally reduces the number of triangles. Also, note that the histogram still shows the peak at one, confirming the aforementioned one-to-one correspondence. \textbf{b}, For the first of the ten rewired patent networks, the temporal decline of the rewired $\mathrm{CD}_5$ mirrors the decline of the original patent network. Since the zero reference patents are preserved by the rewiring algorithm and the majority of the hidden outliers make zero references (Extended Data Fig.~\ref{Extendedfig2}\textbf{a}), this observation provides yet another proof that the hidden outliers are driving the decline. In other words, even upon random citation behaviour, having a certain relative number of zero reference patents with $\mathrm{CD}_{5}=1$ per year induces a decline nearly identical to the one reported in the original manuscript of Park et al.~\citep{park2023papers} (Extended Data Fig.~\ref{Extendedfig2}\textbf{c}). \textbf{c}--\textbf{t}, The equivalent plots for the remaining nine rewiring runs show similar results.
}\label{SupplementalFigure_Patents_rewiring}
\end{figure}

\begin{figure}[ht!]%
\centering
\includegraphics[width=1.0\textwidth]{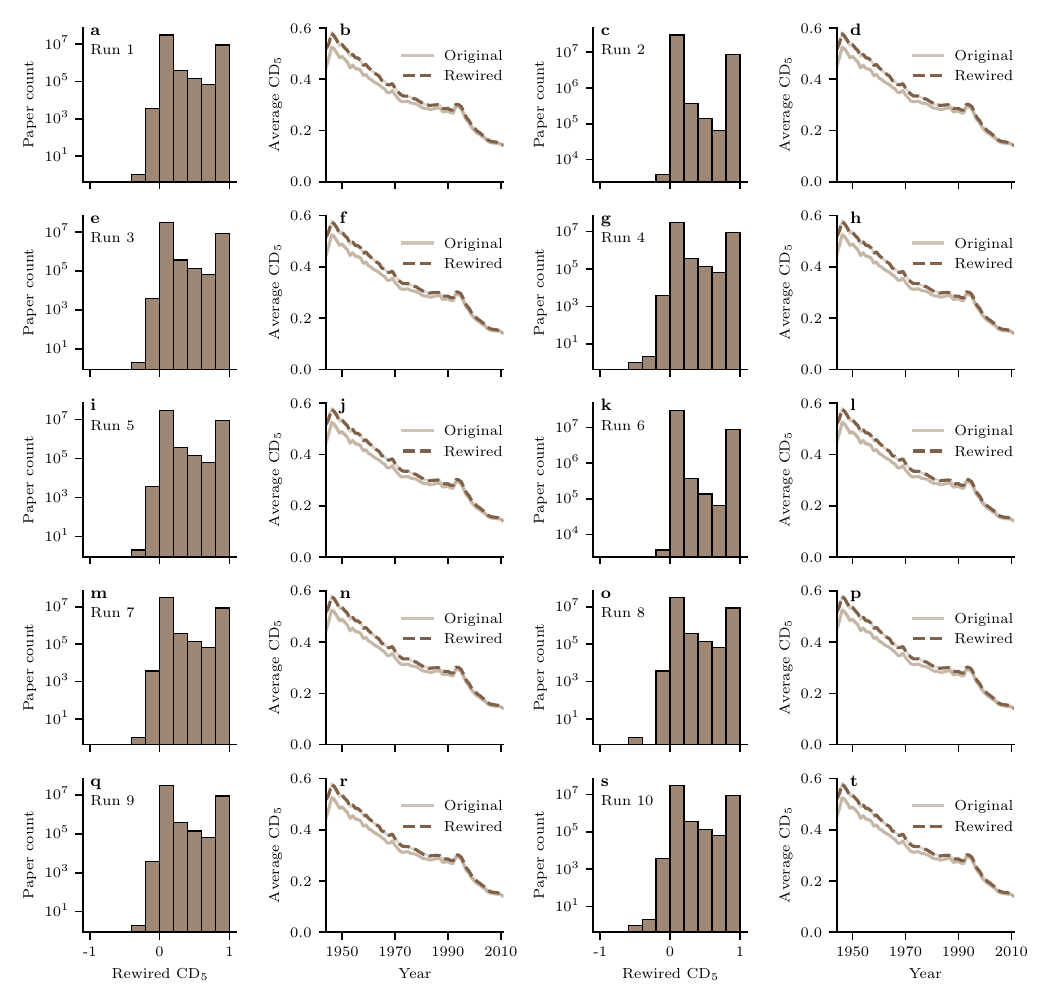}
\vspace{-0.5 cm}
\caption{\textbf{$\vline$ The temporal decline of the $\mathrm{CD}_5$ index for the \textit{SciSciNet} data source is mirrored by random citation behaviour supporting that the hidden outliers are driving the decline.} This figure displays the distribution (panels \textbf{a}, \textbf{c}, \textbf{e}, \textbf{g}, \textbf{i}, \textbf{k}, \textbf{m}, \textbf{o}, \textbf{q}, \textbf{s}) and time average (panels \textbf{b}, \textbf{d}, \textbf{f}, \textbf{h}, \textbf{j}, \textbf{l}, \textbf{n}, \textbf{p}, \textbf{r}, \textbf{t}) of the $\mathrm{CD}_5$ index for ten randomly rewired \textit{SciSciNet} data sources \citep{lin2023sciscinet} (with $39,888,199$ papers), replicating the findings of Supplementary Fig. \ref{SupplementalFigure_Patents_rewiring}. We use the same random rewiring algorithm~\citep{uzzi2013atypical} as Park et al.~\citep{park2023papers}. The shaded bands in the plots correspond to $95 \: \%$ confidence intervals. \textbf{a}, The distribution of the rewired $\mathrm{CD}_5$ for the first of the ten randomly rewired paper networks shows that the algorithm used by Park et al.~\citep{park2023papers} boosts CD index values. The peak at one indicates that random rewiring preserves the zero reference papers. \textbf{b}, For the first of the ten rewired papers networks, the temporal decline of the rewired $\mathrm{CD}_5$ mirrors the decline of the original papers network. Since the zero reference papers are preserved by the rewiring algorithm and the majority of the hidden outliers make zero references (Extended Data Fig.~\ref{Extendedfig2}\textbf{d}), this observation provides yet another proof that the hidden outliers are driving the decline. \textbf{c}--\textbf{t}, The equivalent plots for the remaining nine rewiring runs show similar results.
}\label{SupplementalFigure_SciSciNet_rewiring}
\end{figure}

To evaluate the Monte Carlo simulations, Park et al.~\citep{park2023papers} calculated an average $z$ score among papers or patents published in each year. For an individual paper or patent, the $z$ score is given by:
\begin{equation} \label{definition_z_score}
    \frac{\mathrm{CD}_{\mathrm{observed}}-\mu_\mathrm{rewired}}{\sigma_\mathrm{rewired}}.
\end{equation}
Here, $\mathrm{CD}_{\mathrm{observed}}$ denotes the observed $\mathrm{CD}_5$ index in the unaltered data source, $\mu_\mathrm{rewired}$ denotes the average $\mathrm{CD}_5$ index of the same paper or patent calculated across ten randomly rewired citation networks and $\sigma_\mathrm{rewired}$ denotes the corresponding standard deviation. 
Based on the temporal decline of the average $z$ score (Extended Data Fig. 8\textbf{c,f} in \citep{park2023papers}) the authors concluded ``We find that on average, papers and patents tend to be less disruptive than would be expected by chance, and moreover, the gap between the observed CD index values and those from the randomly rewired networks is increasing over time, which is consistent with our findings of a decline in disruptive science and technology.''
However, the main findings in Park et al.~\citep{park2023papers} are based on the decline of the average $\mathrm{CD}_5$ over time (Fig. 2 in \citep{park2023papers}). To test the robustness of these results against random rewiring, it is therefore logical to also evaluate the rewired $\mathrm{CD}_5$ against time (Fig. \ref{fig2}\textbf{e,f}, Supplementary Fig. \ref{SupplementalFigure_Patents_rewiring}, \ref{SupplementalFigure_SciSciNet_rewiring}, \ref{SupplementalFigure_DBLPv14_rewiring}). These plots unambiguously show that the aforementioned gap is, in fact, narrowing over time. The temporal decrease of the average $z$ score can therefore be attributed to the following phenomenon: the gap between the rewired and observed CD indices shown in Fig. \ref{fig2}\textbf{e,f} corresponds to the (averaged) numerator of the above equation, which indicates that the (averaged) denominator of the above equation, the variance/standard deviation within the ten randomly rewired citation networks, decreases over time (with the caveat that the mean of $\frac{a}{b}$ is of course not exactly equal to the mean of $a$ divided by the mean of $b$). 

\clearpage
\section{~DBLP citation network}
\begin{figure}[ht!]%
\centering
\includegraphics[width=1\textwidth]{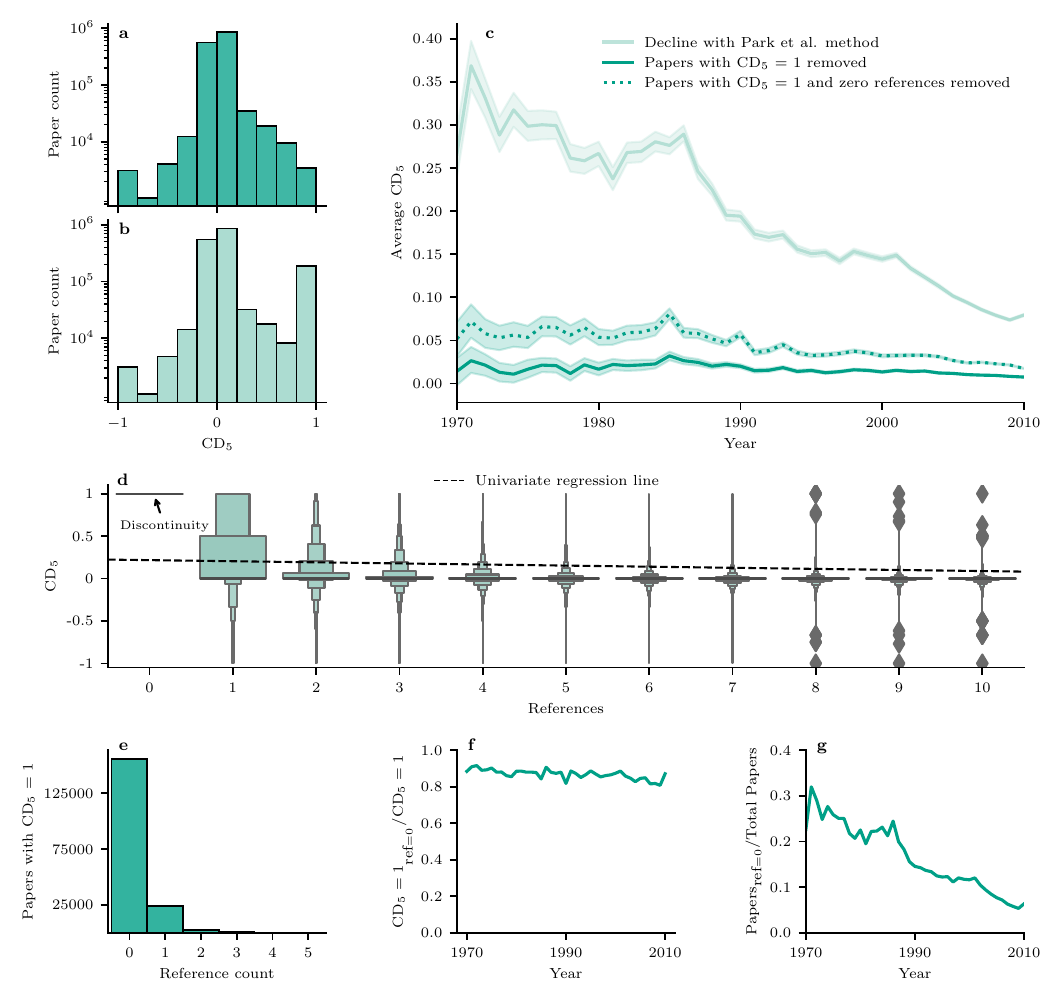}
\vspace{-0.3cm}
\caption{\textbf{$\vline$ Distribution of the $\mathrm{CD}_5$ index with vs without the hidden outliers and its impact on the disruptiveness for the \textit{DBLP-Citation-network V14}.} 
This figure replicates the observation that papers with $\mathrm{CD}_5 = 1$ are driving the decline in disruptive science the \textit{DBLP-Citation-network V14} \citep{tang2008arnetminer}. \textbf{a}, The distribution of the $\mathrm{CD}_5$ index, created using the binwidth parameter in \textit{seaborn 0.11.2}. Here again, the largest data points are hidden. \textbf{b,} The correct histogram of the underlying dataset. A peak at $\mathrm{CD}_5=1$ is revealed, corresponding to $182,398$ additional papers. \textbf{c,} The time evolution of the average $\mathrm{CD}_5$ index. When dropping the outliers with $\mathrm{CD}_5=1$, the decline in disruptiveness is negated. Removing papers with zero references impacts the decline similarly. Moreover, the curve with papers with $\mathrm{CD}_5 = 1$ omitted is the curve corresponding to the histogram~(a). The shaded bands correspond to $95\%$ confidence intervals. \textbf{d,} The distribution of the $\mathrm{CD}_5$ per number of references is shown via letter-value plots which first identify the median, then extend boxes outward, each covering half of the remaining data~\citep{hofmann2017value}. The univariate regression line shows that an ordinary least squared regression fails to capture the discontinuous effect of zero references (Fig. \ref{fig2} \textbf{a,b}). \textbf{e,} The \textit{DBLP-Citation-network V14} contains $182,398$ papers with $\mathrm{CD}_5=1$ between $1970$ and $2010$, of which $85 \: \%$ appear in the database with zero references. \textbf{f,} Within the category of papers with $\mathrm{CD}_5=1$, the relative frequency of papers with zero references is stable between $1970$ and $2010$. \textbf{g,} The relative frequency of papers with $\mathrm{CD}_5$ index exactly equal to one and zero references is decreasing over time, resembling the shape of the top curve shown in panel~(c).}\label{SupplementalFig_DBLPv14}
\end{figure}

In this section, we replicate the same analysis for another independent paper dataset, the \textit{DBLP-Citation-network V14} \citep{tang2008arnetminer}. This data source contains $1,683,086$ papers published between $1970$ and $2010$ in the field of Computer Science.
\begin{itemize}
    \item Fig.~\ref{SupplementalFig_DBLPv14} shows analogous observations for the \textit{DBLP-Citation-network V14} datasets as Fig.~\ref{fig1} by replicating the $\mathrm{CD}_5$ with and without outliers. Since the \textit{DBLP-Citation-network V14} only contains the publication dates of the papers in YYYY format, the $\mathrm{CD}_5$ index is calculated as described in Supplementary Equation \ref{DBLP_cd5}. \\
    \item Fig.~\ref{SupplementalFigure_DBLPv14_rewiring} performs the rewiring analysis for the \textit{DBLP-Citation-network V14}. 
\end{itemize}

\begin{figure}[ht!]%
\centering
\includegraphics[width=1.0\textwidth]{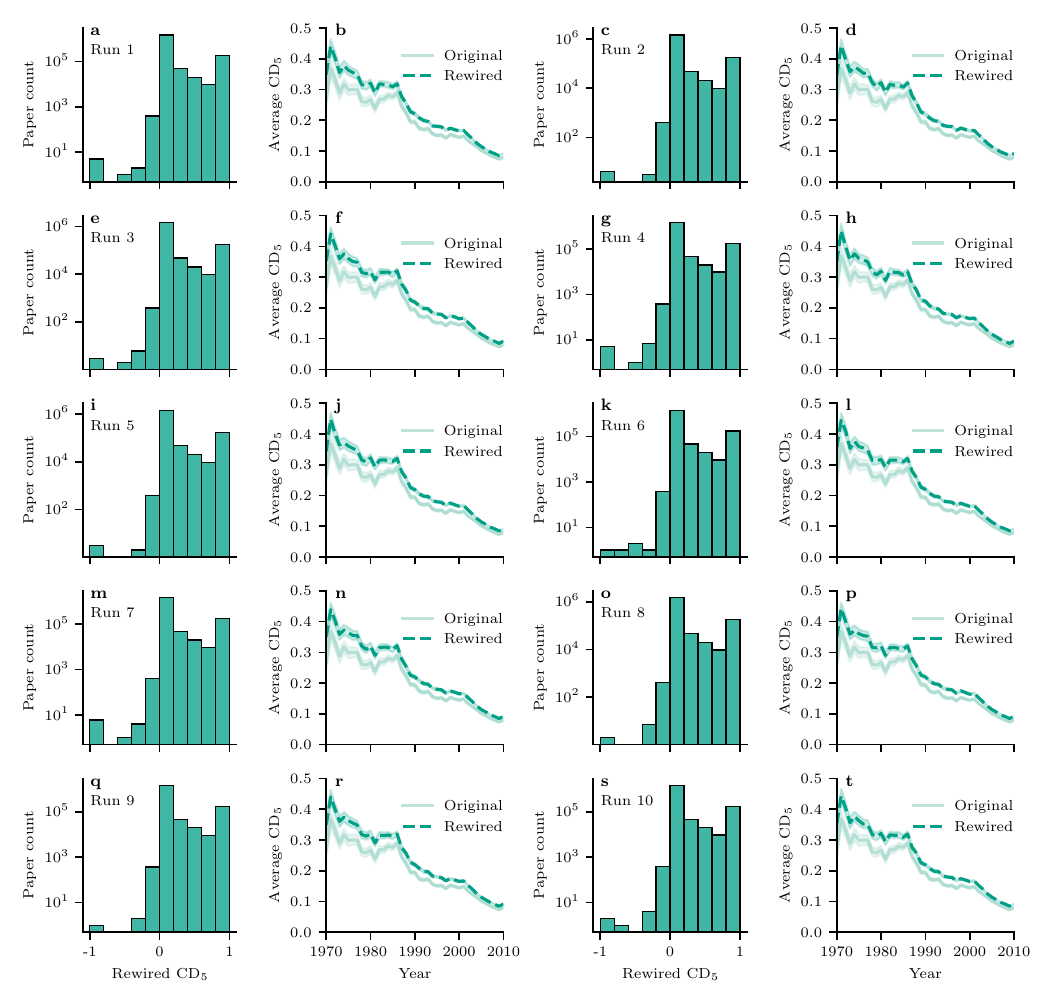}
\vspace{-0.5 cm}
\caption{\textbf{$\vline$ The temporal decline of the $\mathrm{CD}_5$ index for the \textit{DBLP-Citation-network V14} is mirrored by random citation behaviour supporting that the hidden outliers are driving the decline.} This figure displays the distribution (panels \textbf{a}, \textbf{c}, \textbf{e}, \textbf{g}, \textbf{i}, \textbf{k}, \textbf{m}, \textbf{o}, \textbf{q}, \textbf{s}) and time average (panels \textbf{b}, \textbf{d}, \textbf{f}, \textbf{h}, \textbf{j}, \textbf{l}, \textbf{n}, \textbf{p}, \textbf{r}, \textbf{t}) of the $\mathrm{CD}_5$ index for ten randomly rewired \textit{DBLP-Citation-network V14} data sources, replicating Supplementary Fig. \ref{SupplementalFigure_Patents_rewiring} and \ref{SupplementalFigure_SciSciNet_rewiring}. We use the same random rewiring algorithm \citep{uzzi2013atypical} as Park et al.~\citep{park2023papers}. The shaded bands in the plots correspond to $95 \: \%$ confidence intervals. \textbf{a}, The distribution of the rewired $\mathrm{CD}_5$ for the first of the ten randomly rewired paper networks shows that the algorithm used by Park et al.~\citep{park2023papers} boosts CD index values. The peak at one indicates that random rewiring preserves the zero reference papers. \textbf{b}, For the first of the ten rewired papers networks, the temporal decline of the rewired $\mathrm{CD}_5$ mirrors the decline of the original network. Since the zero reference papers are preserved by the rewiring algorithm and the majority of the hidden outliers make zero references (Supplementary Fig.~\ref{SupplementalFig_DBLPv14}\textbf{e}), this observation provides yet another proof that the hidden outliers are driving the decline. \textbf{c}--\textbf{t}, The equivalent plots for the remaining nine rewiring runs show similar results.
}\label{SupplementalFigure_DBLPv14_rewiring}
\end{figure}

\clearpage
\section{~Different forward citation windows}
In this section, we analyse the role of the outliers in the decline for $\mathrm{CD}$ indices with different forward citation windows. We do this for both the \textit{SciSciNet} \citep{lin2023sciscinet} and the \textit{PatentsView} data source.

\begin{figure}[ht]%
\centering
\includegraphics[width=1.0\textwidth]{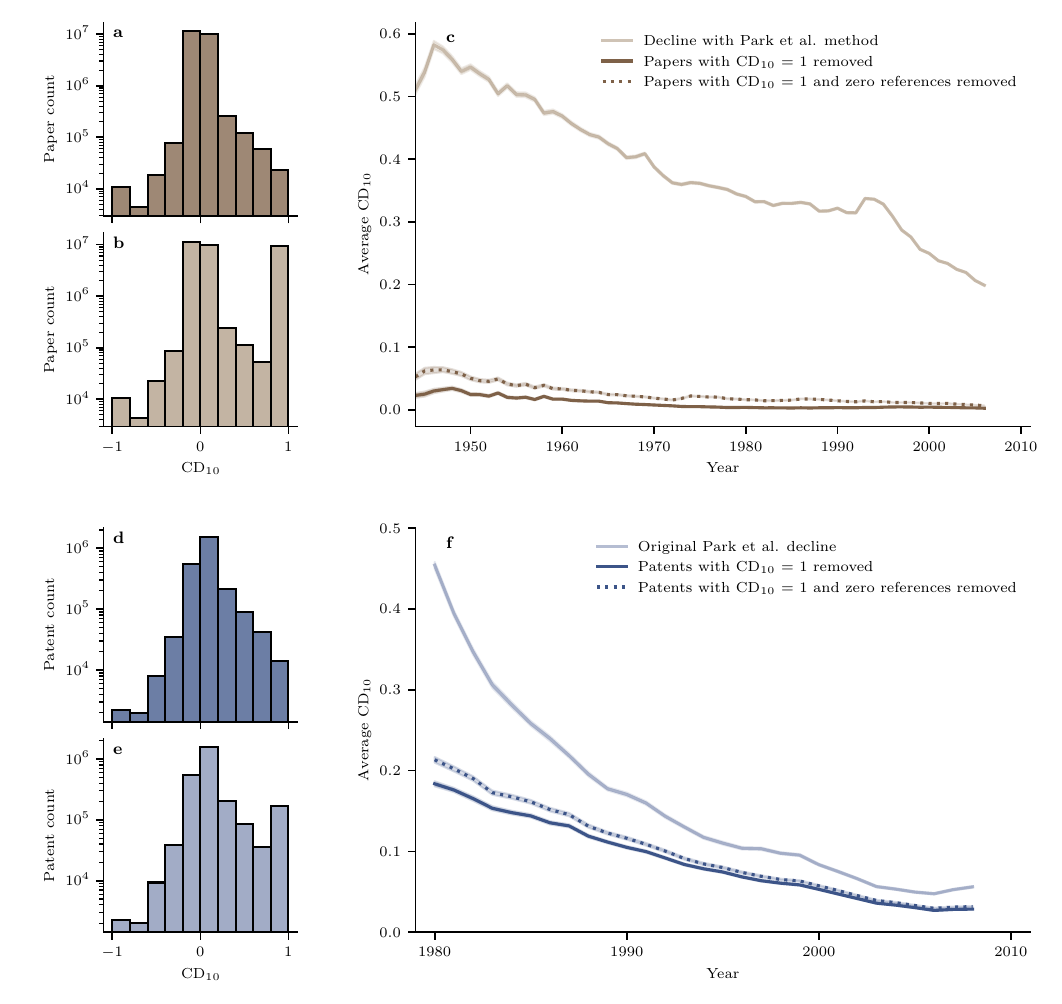}
\vspace{-0.65cm}
\caption{\textbf{$\vline$ Distribution of the $\mathrm{CD}_{10}$ index with vs without the hidden outliers and its impact on the disruptiveness for the \textit{SciSciNet} and the \textit{PatentsView} data source.} 
This figure displays the distribution and time average of the $\mathrm{CD}_{10}$ index (computed over a forward citation window of ten years) for the \textit{SciSciNet} \citep{lin2023sciscinet} data source (with $30,982,865$ papers until 2006) and the \textit{PatentsView} data source (with $2,645,344$ patents until 2008). Importantly, analogously to the $\mathrm{CD}_5$ index, the $\mathrm{CD}_{10}$ index of papers and patents with zero references is either exactly equal to one (if they receive at least one citation within ten years after publication), or remains undefined. \textbf{a}, The distribution of the $\mathrm{CD}_{10}$ index for \textit{SciSciNet}, created using the binwidth parameter in \textit{seaborn 0.11.2}. Here again, the largest data points are hidden. \textbf{b,} The correct histogram of the underlying dataset. A peak at $\mathrm{CD}_{10}=1$ is revealed, corresponding to $9,187,034$ additional papers. \textbf{c,} The time evolution of the average $\mathrm{CD}_{10}$ index. When dropping the outliers with $\mathrm{CD}_{10}=1$, the decline in disruptiveness is negated. We find that $98\%$ of the $\mathrm{CD}_{10} = 1$ papers make zero references, consequently their exclusion impacts the data similarly. Therefore, our claim that papers with a CD index equal to one and zero references are driving the decline in the disruptiveness of scientific knowledge over time is unlikely to be dependent on the size of the forward citation window, which is used to calculate the respective CD index. The shaded bands correspond to $95\:\%$ confidence intervals. \textbf{d--f,} The equivalent plots for \textit{PatentsView} revealing $153,027$ patents with $\mathrm{CD}_{10}=1$. When dropping the outliers with $\mathrm{CD}_{10}=1$, the decline in disruptiveness reduces substantially. We find that $87\:\%$ of the $\mathrm{CD}_{10} = 1$ patents make zero references, consequently their exclusion impacts the decline similarly.}\label{Supplementalfig_cd_10}
\end{figure}

\begin{figure}[ht]%
\centering
\includegraphics[width=1.0\textwidth]{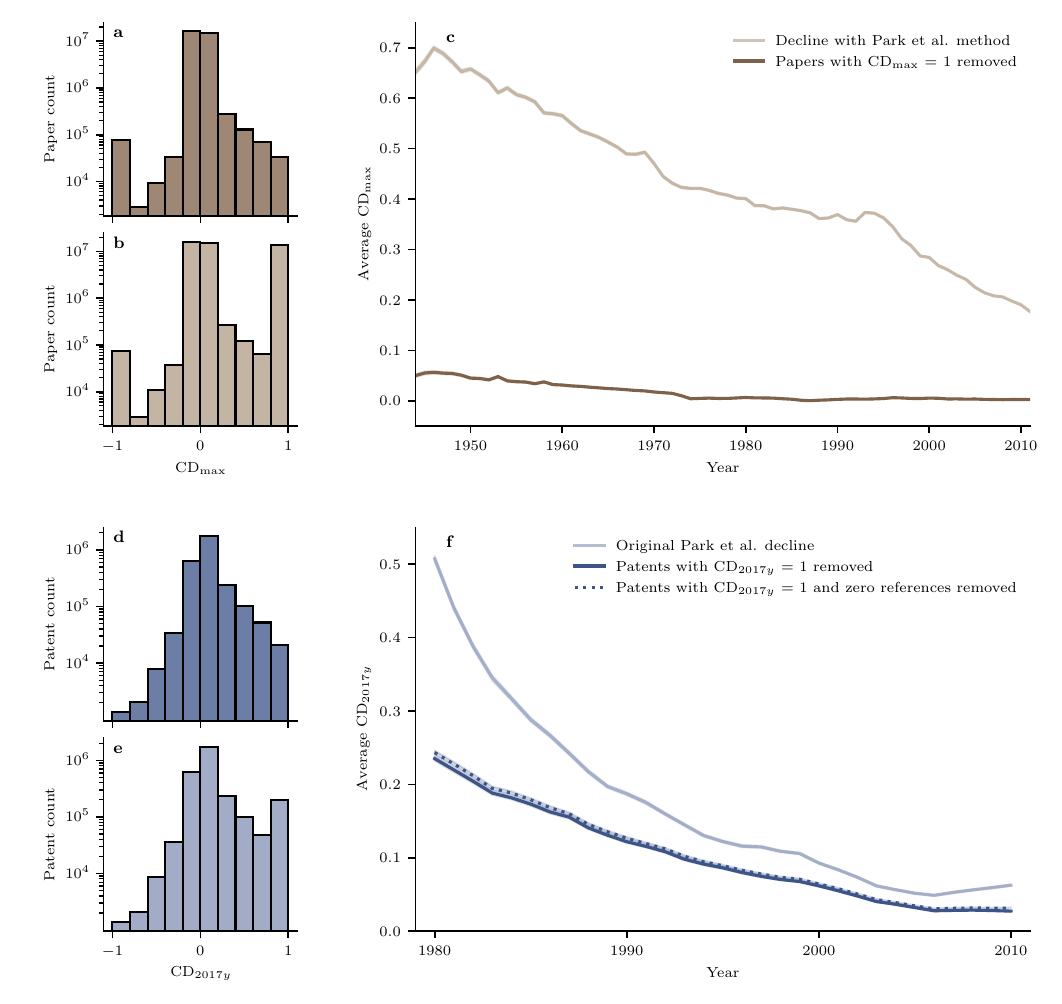}
\vspace{-0.7cm}
\caption{\textbf{$\vline$ Distribution of the $\mathrm{CD}_{\text{max}}$ index with vs without the hidden outliers and its impact on the disruptiveness for the \textit{SciSciNet} and the \textit{PatentsView} data source.} 
This figure displays the distribution and time average of the $\mathrm{CD}_{\text{max}}$ index (computed over a maximum forward citation window) for the \textit{SciSciNet} \citep{lin2023sciscinet} data source (with $45,564,829$ papers) and the \textit{PatentsView} data source (with $3,011,723$ patents). In the notation of Park et al.~\citep{park2023papers}, we have $\mathrm{CD}_{\text{max}}=\mathrm{CD}_{2022y}$ for papers and $\mathrm{CD}_{\text{max}}=\mathrm{CD}_{2017y}$ for patents. Importantly, the $\mathrm{CD}_{\text{max}}$ index of papers and patents with zero references is still either exactly equal to one (if they receive at least one citation), or remains undefined. \textbf{a}, The distribution of the $\mathrm{CD}_{\text{max}}$ index for \textit{SciSciNet}, created using the binwidth parameter in \textit{seaborn 0.11.2}. Here again, the largest data points are hidden. \textbf{b,} The correct histogram of the underlying dataset. A peak at $\mathrm{CD}_{\text{max}}=1$ is revealed, corresponding to $13,864,845$ additional papers. \textbf{c,} The time evolution of the average $\mathrm{CD}_{\text{max}}$ index. When dropping the outliers with $\mathrm{CD}_{\text{max}}=1$, the decline in disruptiveness is negated. We find that all of the $\mathrm{CD}_{\text{max}} = 1$ papers make zero references, consequently their exclusion is not shown separately. The shaded bands correspond to $95\%$ confidence intervals. \textbf{d--f,} The equivalent plots for \textit{PatentsView} revealing $175,190$ patents with $\mathrm{CD}_{2017y}=1$. When dropping the outliers with $\mathrm{CD}_{2017y}=1$, the decline in disruptiveness reduces substantially. We find that $94 \: \%$ of the $\mathrm{CD}_{2017y} = 1$ patents make zero references, consequently their exclusion impacts the decline similarly.}\label{Supplementalfig_cd_max}
\end{figure}

\begin{itemize}
    \item Fig.~\ref{Supplementalfig_cd_10} shows the same analysis as Fig.~\ref{fig1} with the $\mathrm{CD}_{10}$ index for both \textit{SciSciNet} and \textit{PatentsView}. Contrary to the $\mathrm{CD}_5$ index, the $\mathrm{CD}_{10}$ index considers forward citations published within $10$ years after the publication of the focal paper. \\
    \item Fig.~\ref{Supplementalfig_cd_max} shows the same analysis as Fig.~\ref{fig1} with the  $\mathrm{CD}_{\text{max}}$ index for both \textit{SciSciNet} and \textit{PatentsView}.  Contrary to the $\mathrm{CD}_5$ index, the $\mathrm{CD}_{\text{max}}$ index considers all forward citations of a focal paper or patent. For \textit{SciSciNet}, we used the precomputed disruption indices provided by \citep{lin2023sciscinet} for papers with at least one forward citation and one reference. To allow comparison with the Park et al.~\citep{park2023papers} method, we imputed the values with at least one forward citation and zero references to one and the values with zero forward citations and at least one reference to zero.  \\
\end{itemize}

\clearpage
\section{~Normalized $\mathrm{CD}_5$ indices}
\begin{figure}[ht!]%
\centering
\includegraphics[width=1.0\textwidth]{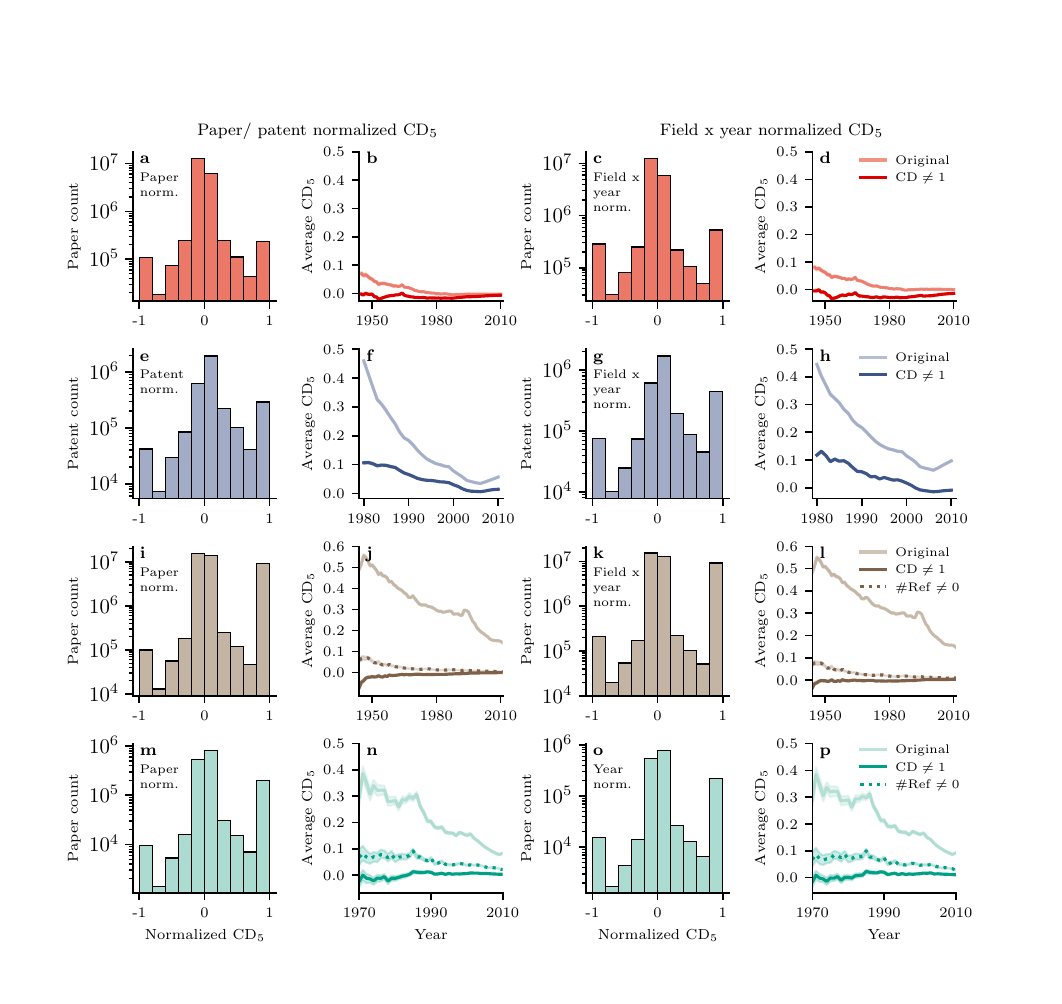}
\vspace{-1.1cm}
\caption{\textbf{$\vline$ Distribution of normalized $\mathrm{CD}_5$ indices and the impact of the hidden outliers on the perceived temporal decline.} This figure shows how the hidden outliers are driving the decline for two normalized CD index variants used by Park et al.~\citep{park2023papers} (Extended Data Fig. 8\textbf{a,d} in \citep{park2023papers}) across four different data sources: \textit{Web of Science}: $19,927,359$ (resp. $19,743,919$) papers (\textbf{a}, \textbf{b}, \textbf{c}, \textbf{d}), \textit{PatentsView}: $3,396,624$ (resp. $3,270,187$) patents, (\textbf{e}, \textbf{f}, \textbf{g}, \textbf{h}), \textit{SciSciNet}: $39,473,940$ (resp. $38,793,453$) papers, (\textbf{i}, \textbf{j}, \textbf{k}, \textbf{l}) and the \textit{DBLP-Citation-network V14}: $1,651,398$ (resp. $1,623,775$) papers,  (\textbf{m}, \textbf{n}, \textbf{o}, \textbf{p}). Counts differ between the two normalized $\mathrm{CD}$ indices since they can be undefined for different papers and patents. The normalized variants adjust the term $N_R$ (resp. $N_k$ in the notation of Park et al.~\citep{park2023papers}; see also Supplementary equation \ref{definition_cd_2}) in the definition of the CD index, i.e. they modify the part of the definition that refers to follow-up work that does not cite the focal paper or patent itself but at least one of its references. A detailed definition of both variants can be found in the introduction of the present section. It is worth noting that for both normalized variants, papers and patents that make zero references have a value that either is exactly equal to one (if they are cited at least once) or remains undefined. Panels \textbf{a}, \textbf{e}, \textbf{i}, \textbf{m} reveal a peak at one for the paper (resp. patent) normalized CD index across all aforementioned data sources. Panels \textbf{b}, \textbf{f}, \textbf{j}, \textbf{n} show the time evolution of the average paper (resp. patent) normalized $\mathrm{CD}_5$ index for papers. When dropping the hidden outliers with normalized CD index exactly equal to one, the decline in disruptiveness completely disappears for the paper datasets and substantially reduces for the patent dataset. For the \textit{SciSciNet} data source and the \textit{DBLP-Citation-network V14}, we had access to sufficient metadata to also exclude papers that make zero references similarly impacting the decline. Shaded bands correspond to $95 \%$ confidence intervals. The remaining panels show the equivalent plots for the field x year normalized CD index. Being specific to the field of computer science, we show the year normalized CD index for \textit{DBLP-Citation-network V14}.}\label{Supplementalfig_disruption_indices}
\end{figure}

In this section, we analyse the distribution of the normalized $\mathrm{CD}_5$ indices and the impact of the hidden outliers on the perceived temporal decline. We do this for four data sources: \textit{Web of Science}, \textit{PatentsView}, \textit{SciSciNet} \citep{lin2023sciscinet} and the \textit{DBLP-Citation-network V14} \citep{tang2008arnetminer}. The two normalized CD index variants are the ones used by Park et al.~\citep{park2023papers} (Extended Data Fig. 8\textbf{a,d} in \citep{park2023papers}). Both variants adjust the term $N_R$ (resp. $N_k$ in the notation of Park et al.~\citep{park2023papers}) in the definition of the CD index, i.e. they modify the part of the definition that refers to follow-up work that does not cite the focal paper or patent itself but at least one of its references (Supplementary Equation \ref{definition_cd_1} and \ref{definition_cd_2}). 
\begin{itemize}
    \item Paper (resp. patent) normalized CD index: replace $N_R$ with $\mathrm{max}(N_R-\#\text{ref},\: 0)$. \\
    \item Field x year normalized CD index: replace $N_R$ with $\mathrm{max}(N_R-\overline{\#\text{ref}}(\text{field, year}),\: 0)$. 
\end{itemize}
 Here, $\#\text{ref}$ denotes the number of references of the focal paper (resp. patent), $\overline{\#\text{ref}}(\text{field, year})$ denotes the average number of references per year and field, and $\mathrm{max}(\cdot, \cdot)$ denotes the maximum between two values. We show our analysis in Fig.~\ref{Supplementalfig_disruption_indices}.

\clearpage
\section{~Random paper and patent samples}
This section contains tables with additional information on the $100$ randomly drawn papers and patents from the sample with zero references and at least one forward citation within five years after publication, which results in $\mathrm{CD}_5=1$.
\\
\begin{table}[b!]
\centering
\caption{Summary statistics of the randomly drawn papers and patents with $\mathrm{CD}_5=1$ and zero references.}
\label{tab:summary_sample}
\begin{tabular}{cccccccccc}
\toprule
\textbf{Papers} & Ref. & Zero ref. & Conference & Technical & List of & News \\ 
& in PDF & in PDF & & report & abstracts & \\
\addlinespace
\addlinespace
& $93\%$ & $7\%$ & $1\%$ & $1\%$ & $4\%$ & $1\%$ \\
\addlinespace
\midrule
\addlinespace
\textbf{Patents} & Ref. & Zero ref. & US & US & A1 & E & Foreign & Other \\ 
& in PDF & in PDF & Pre-1976 & Post-1976 \\
\addlinespace
\addlinespace
& $98\%$ & $2\%$ & $49\%$  & $1\%$ & $16\%$ & $1\%$ & $48\%$ & $39\%$ \\
\addlinespace
\bottomrule
\end{tabular}
{\footnotesize 
Note: 
This table contains the summary statistics of $100$ randomly drawn papers and patents from the sample with zero references and at least one forward citation within five years after publication, which results in $\mathrm{CD}_5=1$. For \textit{SciSciNet}, we draw papers from the \textit{pandas DataFrame} with zero references and at least one forward citation using a random seed equal to zero. Since we do not have access to all the PDF files, we have to draw $238$ papers until we obtain $100$ accessible PDFs. The random sample for patents is drawn from the \textit{pandas DataFrame} provided by Park et al.~\citep{park2023papers} using a random seed equal to zero.
We manually verified the PDF files (see Supplementary Tables \ref{tab:paper_details}, \ref{tab:patent_details}) for references and find that $93\%$ of papers and $98\%$ of patents make at least one reference.
For papers, we have a more detailed view on the type of the seven documents that do not make any references to prior work. We find four papers which are part of a list of abstracts. Further, there is a conference paper which mentions prior work but does not make explicit references, one Nature news article, and one technical report in the transportation research record.
For patents, we have a more detailed view on which types of references are missing and report the percentage of patents which contain at least one reference to a specific category. Note that each patent may contain references to multiple types of references. We find $49\%$ of patents contain references that are missing due to truncation caused by pre-processing the data (pre-1976) \citep{macher2023illusive}, $16\%$ due to not accounting for patent applications since the passage of the Inventor Protection Act of 1999 (A1) \citep{macher2023illusive}, and $48\%$ and $39\%$ due to not counting foreign patents and other publications, respectively. Only one patent (id 6552498) contains a reference to a US patent, which seems to be a bibliometric error, and one patent contains a reference to a reissue patent (E).
Finally, we find that these papers and patents have a median of $17.5$ and $5$ references which corresponds to the median of $9$ and $5$ references for the full sample, respectively. We compute the median for papers only for PDF files where we are able to count the exact number of references, i.e. by excluding the ``1+'' (see Supplementary Table \ref{tab:paper_details} for details on the occurrences of ``1+'' in the case of only partial accessibility of the PDF file or references in footnotes).
}
\end{table}

\clearpage
\begin{landscape}
\begin{longtable}{|l|p{0.8cm}|p{0.8cm}|p{1cm}|p{2.1cm}|p{1.6cm}|p{1cm}|p{1.9cm}|p{1.6cm}|p{1.5cm}|p{0.9cm}|}
\caption{Paper details.}\label{tab:paper_details} \\
\hline
\textbf{Paper ID} & \textbf{Year} & \textbf{Link} & \textbf{Open access} & \textbf{\#References in PDF} & \textbf{PDF available} & \textbf{Foot-notes} & \textbf{Conference} & \textbf{Technical report} & \textbf{List of abstracts} & \textbf{News} \\
\hline
\endfirsthead

\hline
\textbf{Paper ID} & \textbf{Year} & \textbf{Link} & \textbf{Open access} & \textbf{\#References in PDF} & \textbf{PDF available} & \textbf{Foot-notes} & \textbf{Conference} & \textbf{Technical report} & \textbf{List of abstracts} & \textbf{News} \\
\hline
\endhead

\hline
\endfoot

27985486 & 2008 & \href{https://www.proquest.com/openview/33ac2f9f78cf4cf52c0114927b4ca956}{Link} & no & 27 &  &  & & & & \\ \hline
53937866 & 1993 & \href{https://www.researchgate.net/profile/Joseph-Locicero/publication/14931560_Credentialing_issues_and_complications_of_video-assisted_thoracic_surgery/links/0912f50196c313670c000000/Credentialing-issues-and-complications-of-video-assisted-thoracic-surgery.pdf}{Link} & yes & 6 &  &  & & & & \\ \hline
98255911 & 1998 & \href{https://wow-ortho.com/wp-content/uploads/2022/05/Connecticut_Intrusion_Arch.pdf}{Link} & yes & 13 &  &  & & & & \\ \hline
150390299 & 1994 & \href{https://books.google.be/books?hl=en&lr=&id=9RAnJzaiqvMC&oi=fnd&pg=PA199&dq=Computer+conferencing+and+the+new+Europe}{Link} & yes & 1+ & partial & no & & & & \\ \hline
180437480 & 2000 & \href{https://journals.lww.com/acsm-essr/Abstract/2000/28040/Muscle_Glucose_Transporter__GLUT_4__Gene.2.aspx}{Link} & yes & 15 &  &  & & & & \\ \hline

189909736 & 2006 & \href{https://www.ptfarm.pl/pub/File/Acta_Poloniae/2006/1/63.pdf}{Link} & yes & 18 &  &  & & & & \\ \hline
288606890 & 1963 & \href{https://www.ncbi.nlm.nih.gov/pmc/articles/PMC1949674/pdf/amjpathol00316-0042.pdf}{Link} & yes & 6 &  &  & & & & \\ \hline
574253104 & 2006 & \href{https://www.torrossa.com/it/resources/an/4815976}{Link} & yes & 1+ & partial & yes & & & & \\ \hline
575352733 & 2011 & \href{https://books.google.be/books?hl=en&lr=&id=N4IF2p4w7uwC&oi=fnd&pg=PA361&dq=Machine+Ethics:+Piagetian+Roboethics+via+Category+Theory}{Link} & yes & 1+ & partial & no & & & & \\ \hline
585396547 & 2000 & \href{https://books.google.be/books?hl=en&lr=&id=hkBBDgAAQBAJ&oi=fnd&pg=PT6&dq=The+political+ecologist+wells}{Link} & yes & 1+ & partial & no & & & & \\ \hline
585713240 & 1991 & \href{https://onlinepubs.trb.org/Onlinepubs/trr/1991/1311/1311-009.pdf}{Link} & yes & 0 & & & no & yes & no & no \\ \hline
639863130 & 2007 & \href{https://journals.lww.com/joacp/_layouts/15/oaks.journals/downloadpdf.aspx?an=01607275-200723040-00009}{Link} & yes & 11 &  &  & & & & \\ \hline
952410954 & 2011 & \href{https://repositorio.unicamp.br/acervo/detalhe/847119}{Link} & yes & 68 &  &  & & & & \\ \hline
1411236576 & 1972 & \href{https://www.tandfonline.com/doi/pdf/10.1080/00021369.1972.10860287}{Link} & yes & 9 &  &  & & & & \\ \hline
1493574560 & 1958 & \href{https://repository.rothamsted.ac.uk/download/da923ac105983581f7731b3062a119a7e690931c120a16aad0550dd9c527e8fe/10768968/The%20Design%20and%20Analysis.pdf}{Link} & yes & 18 &  &  & & & & \\ \hline
1495110460 & 2003 & \href{https://dspace.mit.edu/bitstream/handle/1721.1/8034/52724850-MIT.pdf}{Link} & yes & 17 &  &  & & & & \\ \hline
1522126398 & 1977 & \href{https://journals.aai.org/jimmunol/article/118/6/2015/10101/Two-Migration-Inhibitory-Factors-with-Different}{Link} & no & 12 &  &  & & & & \\ \hline
1536891849 & 1976 & \href{https://link.springer.com/content/pdf/10.1007/3-540-07804-5_31.pdf}{Link} & yes & 0 &  &  & yes & no & no & no \\ \hline

1557088112 & 2002 & \href{https://books.google.be/books?hl=en&lr=&id=KBaFchM0dowC&oi=fnd&pg=PR9&dq=DAMNED+FOR+THEIR+DIFFERENCE:+THE+CULTURAL+CONSTRUCTION+OF+DEAF+PEOPLE+AS+DISABLED:+A+SOCIOLOGICAL+HISTORY}{Link} & yes & 1+ & partial & yes & & & & \\ \hline
1575011135 & 2009 & \href{https://books.google.be/books?hl=en&lr=&id=rOVtCQAAQBAJ&oi=fnd&pg=PR2&dq=Theorem+proving+in+higher+order+logics+:+22nd+International+Conference,+TPHOLs+2009,+Munich,+Germany,+August+17-20,+2009+:+proceedings}{Link} & yes & 1+ & full & no & & & & \\ \hline
1633088188 & 1991 & \href{https://books.google.be/books?hl=en&lr=&id=0-Lev9BQaXwC&oi=fnd&pg=PR9&dq=Warped+Disks+and+Inclined+Rings+around+Galaxies}{Link} & yes & 1+ & partial & no & & & & \\ \hline
1657310252 & 1987 & \href{https://www.nature.com/articles/329190a0.pdf}{Link} & yes & 0 &  &  & no & no & no & yes \\ \hline
1813921254 & 1975 & \href{https://www.jstor.org/stable/pdf/2662426.pdf}{Link} & yes & 1+ & partial & yes & & & & \\ \hline

\multicolumn{11}{|p{16.5cm}|}{\footnotesize Note: This table continues on the next three pages and contains the details of $100$ randomly drawn papers from the sample with zero references and at least one forward citation, which results in $\mathrm{CD}_5=1$. The random sample is drawn from the \textit{pandas DataFrame} with zero references and at least one forward citation using a random seed equal to zero. Since we do not have access to all the PDF files, we have to draw $238$ papers until we obtain $100$ accessible PDFs.} \\

\newpage
1818446669 & 1990 & \href{https://jnm.snmjournals.org/content/jnumed/31/4/519.full.pdf}{Link} & yes & 25 &  &  & & & & \\ \hline
1856184614 & 1969 & \href{https://www.researchgate.net/profile/Ruben-Seisyan/publication/239664806_Diamagnetic_excitons_in_semiconductors/links/574d86c308aec988526b6ac3/Diamagnetic-excitons-in-semiconductors.pdf}{Link} & yes & 40 &  &  & & & & \\ \hline
1968249834 & 1995 & \href{https://rupress.org/jgp/article-pdf/105/3/421/1768215/421.pdf}{Link} & yes & 53 &  &  & & & & \\ \hline
1968371615 & 1990 & \href{https://hal.science/hal-00885278/document}{Link} & yes & 14 &  &  & & & & \\ \hline
1970177916 & 1993 & \href{https://www.sciencedirect.com/science/article/pii/S0003269783710870}{Link} & no & 30 &  &  & & & & \\ \hline
1971682916 & 1996 & \href{https://pubs.aip.org/asa/jasa/article/100/4_Supplement/2611/487239/Occurrence-of-blue-and-fin-whales-calls-in-the}{Link} & yes & 1+ & partial & no & & & & \\ \hline
1975852712 & 1988 & \href{https://pubs.aip.org/avs/jva/article-abstract/6/3/726/248291/Percolative-aspects-of-nonequilibrium-adlayer}{Link} & no & 25 &  &  & & & & \\ \hline
1977359555 & 2006 & \href{https://ps.psychiatryonline.org/doi/epdf/10.1176/ps.2006.57.12.1758}{Link} & yes & 31 &  &  & & & & \\ \hline
1978870643 & 2011 & \href{https://koreascience.kr/article/JAKO201112049744257.pdf}{Link} & yes & 28 &  &  & & & & \\ \hline
1982053324 & 1995 & \href{https://pubs.acs.org/doi/pdf/10.1021/ic00123a028}{Link} & yes & 1+ & partial & yes & & & & \\ \hline
1985373775 & 1996 & \href{https://onlinelibrary.wiley.com/doi/pdf/10.1111/j.1360-0443.1996.tb02276.x}{Link} & yes & 5 &  &  & & & & \\ \hline
1989942175 & 1990 & \href{https://journals.lww.com/transplantjournal/abstract/1990/10000/successful_use_of_rifampicin_in_the_treatment_of.14.aspx}{Link} & yes & 6 &  &  & & & & \\ \hline
1991296186 & 1993 & \href{https://journals.lww.com/plasreconsurg/Abstract/1993/09000/The_Anatomy_of_Cupid_s_Bow_in_Normal_and_Cleft_Lip.1.aspx}{Link} & yes & 39 &  &  & & & & \\ \hline
1996355016 & 1999 & \href{https://scholar.google.com/scholar?hl=en&as_sdt=0%2C5&q=Making+%28It%29+Public+Jodi+Dean&btnG=}{Link} & yes & 1+ & full & yes & & & & \\ \hline
2007371250 & 2004 & \href{https://www.sciencedirect.com/science/article/abs/pii/S0196064404009758}{Link} & no & 0 &  &  & no & no & yes & no \\ \hline
2008609890 & 1991 & \href{https://www.sciencedirect.com/science/article/pii/073510979191971G}{Link} & no & 0 &  &  & no & no & yes & no \\ \hline
2010293280 & 1990 & \href{https://articles.adsabs.harvard.edu/pdf/1990ApJS...73..199R}{Link} & yes & 42 &  &  & & & & \\ \hline
2023511737 & 1999 & \href{https://journals.aps.org/pra/pdf/10.1103/PhysRevA.60.3000}{Link} & no & 15 &  &  & & & & \\ \hline
2024560885 & 1975 & \href{https://journals.lww.com/academicmedicine/_layouts/15/oaks.journals/downloadpdf.aspx?an=00004999-197504000-00018}{Link} & yes & 3 &  &  & & & & \\ \hline
2027584436 & 2006 & \href{https://www.thieme-connect.com/products/ejournals/html/10.1055/s-2006-923917}{Link} & no & 13 &  &  & & & & \\ \hline
2040840436 & 2006 & \href{https://www.scielo.br/j/jped/a/8dDfNdvjFRq33TYGbfQvFTk/?lang=pt}{Link} & yes & 54 &  &  & & & & \\ \hline
2042086152 & 2010 & \href{https://www.ncbi.nlm.nih.gov/pmc/articles/PMC2868642/}{Link} & yes & 1 &  &  & & & & \\ \hline
2050466013 & 1999 & \href{https://onlinelibrary.wiley.com/doi/pdf/10.1111/1467-8373.00088}{Link} & no & 31 &  &  & & & & \\ \hline
2057336774 & 1993 & \href{https://www.persee.fr/doc/asie_0766-1177_1993_num_7_1_1062}{Link} & yes & 1+ & partial & yes & & & & \\ \hline
2060753453 & 1999 & \href{https://downloads.hindawi.com/archive/1999/517147.pdf}{Link} & yes & 18 &  &  & & & & \\ \hline
2063676441 & 2006 & \href{https://www.arca.fiocruz.br/bitstream/handle/icict/18992/Ensaio_minhoca_Cristina%20Sissino_2006.pdf?sequence=2&isAllowed=y}{Link} & yes & 6 &  &  & & & & \\ \hline
2069640366 & 1992 & \href{https://www.tandfonline.com/doi/pdf/10.1271/bbb.56.1378}{Link} & yes & 22 &  &  & & & & \\ \hline
2081547537 & 1969 & \href{https://karger.com/aan/article-abstract/72/3/357/5565/Histochemical-studies-on-the-early-development-of?redirectedFrom=PDF}{Link} & yes & 1+ & partial & no & & & & \\ \hline
2083436642 & 2010 & \href{https://koreascience.kr/article/JAKO201017337336724.pdf}{Link} & yes & 26 &  &  & & & & \\ \hline
2083638940 & 1991 & \href{https://www.jstor.org/stable/pdf/43447363.pdf}{Link} & yes & 1+ & partial & yes & & & & \\ \hline
2086255442 & 2008 & \href{https://www.jstage.jst.go.jp/article/suisan/74/5/74_5_906/_pdf/-char/en}{Link} & yes & 1 &  &  & & & & \\ \hline
2090072051 & 1991 & \href{https://journals.aps.org/prl/pdf/10.1103/PhysRevLett.66.1259}{Link} & no & 19 &  &  & & & & \\ \hline
2117751013 & 2004 & \href{https://www.cambridge.org/core/services/aop-cambridge-core/content/view/73B80BD9832219087525BA8B5D49DE97/S104161020400081Xa.pdf/div-class-title-cognition-behavior-and-the-frontal-lobes-div.pdf}{Link} & yes & 35 &  &  & & & & \\ \hline
2123154878 & 2008 & \href{https://www.sciencedirect.com/science/article/abs/pii/S019606440800334X}{Link} & no & 0 &  &  & no & no & yes & no \\ \hline
2123761965 & 1968 & \href{https://books.google.be/books?hl=en&lr=&id=kOd8CgAAQBAJ&oi=fnd&pg=PT14&dq=The+History+Of+Scepticism+From+Erasmus+To+Descartes}{Link} & yes & 1+ & partial & yes & & & & \\ \hline
2139136071 & 2011 & \href{https://www.cmaj.ca/content/cmaj/183/7/E385.full.pdf}{Link} & yes & 1+ & full & yes & & & & \\ \hline
2313058271 & 1973 & \href{https://www.neurology.org/doi/pdf/10.1212/WNL.23.12.1257}{Link} & no & 41 &  &  & & & & \\ \hline
2317934052 & 1949 & \href{https://www.jstor.org/stable/pdf/2549415.pdf}{Link} & yes & 1+ & partial & yes & & & & \\ \hline
2328439434 & 1969 & \href{https://www.jstor.org/stable/pdf/1249398.pdf}{Link} & no & 14 &  &  & & & & \\ \hline
2347015509 & 2009 & \href{https://ieeexplore.ieee.org/stamp/stamp.jsp?arnumber=4895361}{Link} & no & 33 &  &  & & & & \\ \hline
2389311988 & 2010 & \href{https://zgnyqx.ieda.org.cn/EN/abstract/abstract622.shtml}{Link} & yes & 32 &  &  & & & & \\ \hline
2405115874 & 1986 & \href{https://www.ncbi.nlm.nih.gov/pmc/articles/PMC1146972/pdf/biochemj00276-0246.pdf}{Link} & yes & 24 &  &  & & & & \\ \hline
2410752987 & 1989 & \href{https://www.researchgate.net/profile/Robert-Rej/publication/20469142_Aminotransferase_in_disease/links/0912f50ca10b7b23e5000000/Aminotransferase-in-disease.pdf}{Link} & yes & 182 &  &  & & & & \\ \hline
2412560019 & 1999 & \href{https://www.sciencedirect.com/science/article/pii/S0362028X22038996}{Link} & yes & 15 &  &  & & & & \\ \hline
2412811714 & 2000 & \href{https://www.researchgate.net/profile/Alev-Alacam/publication/12127038_Microleakage_of_light-cured_resin_and_resin-modified_glass-ionomer_dentin_bonding_agents_applied_with_co-cure_vs_pre-cure_technique/links/6450bc014af788735250648b/Microleakage-of-light-cured-resin-and-resin-modified-glass-ionomer-dentin-bonding-agents-applied-with-co-cure-vs-pre-cure-technique.pdf#page=61}{Link} & yes & 32 &  &  & & & & \\ \hline
2418732624 & 1991 & \href{https://portlandpress.com/biochemsoctrans/article-abstract/19/3/278S/82255/Expression-of-drug-resistance-in-Chinese-hamster}{Link} & yes & 6 &  &  & & & & \\ \hline
2418744057 & 1987 & \href{https://books.google.be/books?hl=en&lr=&id=eyJxS6FTJqMC&oi=fnd&pg=PA689&dq=High-dose+i.v.+thiotepa+and+cryopreserved+autologous+bone+marrow+transplantation+for+therapy+of+refractory+cancer.#v=onepage&q=High-dose%20i.v.%20thiotepa%20and%20cryopreserved%20autologous%20bone%20marrow%20transplantation%20for%20therapy%20of%20refractory%20cancer.&f=false}{Link} & yes & 61 &  &  & & & & \\ \hline
2419118622 & 2002 & \href{https://www.ima.org.il/filesupload/IMAJ/0/56/28247.pdf}{Link} & yes & 5 &  &  & & & & \\ \hline
2461652553 & 1997 & \href{https://www.degruyter.com/document/doi/10.1515/tsd-1997-340609/html}{Link} & no & 51 &  &  & & & & \\ \hline
2484809223 & 2006 & \href{https://books.google.be/books?hl=en&lr=&id=RymGgxN3zD4C&oi=fnd&pg=PP13&dq=The+Handbook+of+Marketing+Research}{Link} & yes & 1+ & partial & yes & & & & \\ \hline
2490486294 & 1974 & \href{https://www.researchgate.net/profile/Rolf-Steiger/publication/307789887_Stereoisomerism_of_Meso_Substituted_Carbocyanines_and_Its_Influence_on_J-aggregate_Formation/links/59734a49a6fdcc8348829c14/Stereoisomerism-of-Meso-Substituted-Carbocyanines-and-Its-Influence-on-J-aggregate-Formation.pdf}{Link} & yes & 36 &  &  & & & & \\ \hline
2503700092 & 1999 & \href{https://books.google.be/books?hl=en&lr=&id=pv-3CgAAQBAJ&oi=fnd&pg=PT228&dq=The+Role+of+Nitric+Oxide+in+Cerebrovascular+Regulation+and+Stroke}{Link} & yes & 147 &  &  & & & & \\ \hline
2507383598 & 1988 & \href{https://books.google.be/books?hl=en&lr=&id=49oGjgz8riQC&oi=fnd&pg=PA21&dq=Biologic+chemistry+of+chromium}{Link} & yes & 1+ & partial & no & & & & \\ \hline
2748776472 & 2006 & \href{https://www.researchgate.net/profile/Luiz-Fernandes-2/publication/323231873_DIAGENESE_DOS_ARENITOS_DO_GRUPO_ITARARE_PERMOCARBONIFERO_BACIA_DO_PARANA/links/5aeb5d800f7e9b01d3e054fc/DIAGENESE-DOS-ARENITOS-DO-GRUPO-ITARARE-PERMOCARBONIFERO-BACIA-DO-PARANA.pdf}{Link} & yes & 40 &  &  & & & & \\ \hline
2886101548 & 2006 & \href{https://www.ahajournals.org/doi/full/10.1161/CIRCULATIONAHA.105.612309}{Link} & yes & 5 &  &  & & & & \\ \hline
3103741798 & 2002 & \href{https://iopscience.iop.org/article/10.1086/341552/pdf}{Link} & yes & 37 &  &  & & & & \\ \hline
3168256524 & 1990 & \href{https://articles.adsabs.harvard.edu/cgi-bin/nph-iarticle_query?bibcode=1990BAAS...22.1041T&db_key=AST&page_ind=0&data_type=GIF&type=SCREEN_VIEW&classic=YES}{Link} & yes & 0 &  &  & no & no & yes & no \\ \hline
3188737630 & 2009 & \href{https://books.google.be/books?hl=en&lr=&id=Mj58AgAAQBAJ&oi=fnd&pg=PA8&dq=Concepts+of+flagships}{Link} & yes & 1+ & partial & no & & & & \\ \hline
75971907 & 1986 & \href{https://academic.oup.com/clinchem/article-abstract/32/6/1246/5652824}{Link} & yes & 2 &  &  & & & & \\ \hline
128280131 & 1990 & \href{https://www.thieme-connect.de/products/ejournals/abstract/10.1055/s-2007-998141}{Link} & yes & 1+ & partial & no & & & & \\ \hline
235473009 & 2010 & \href{https://www.researchgate.net/profile/Aaron-Perzanowski/publication/49249400_Unbranding_Confusion_Deception/links/54b7d0aa0cf28faced6082c2/Unbranding-Confusion-Deception.pdf}{Link} & yes & 1+ & full & yes & & & & \\ \hline
257670102 & 2009 & \href{https://www.researchgate.net/profile/Jinlan-Zhang-2/publication/26256064_Development_of_a_Rapid_Resolution_Liquid_Chromatographic_Method_for_Simultaneous_Analysis_of_Four_Alkaloids_in_Rhizoma_coptidis_Under_Different_Cultivation_Conditions/links/575696f208ae155a87bb2d47/Development-of-a-Rapid-Resolution-Liquid-Chromatographic-Method-for-Simultaneous-Analysis-of-Four-Alkaloids-in-Rhizoma-coptidis-Under-Different-Cultivation-Conditions.pdf}{Link} & yes & 17 &  &  & & & & \\ \hline
288441233 & 1994 & \href{https://www.ncbi.nlm.nih.gov/pmc/articles/PMC1022626/pdf/westjmed00062-0066a.pdf}{Link} & yes & 3 &  &  & & & & \\ \hline
649906705 & 1978 & \href{https://books.google.be/books?hl=en&lr=&id=2b155zcw-GEC&oi=fnd&pg=PP1&dq=Chemistry+and+action+of+herbicide+antidotes}{Link} & yes & 1+ & partial & no & & & & \\ \hline
1004243738 & 1981 & \href{https://europepmc.org/backend/ptpmcrender.fcgi?accid=PMC2396084&blobtype=pdf}{Link} & yes & 3 &  &  & & & & \\ \hline
1488058573 & 1996 & \href{https://www.researchgate.net/profile/Martin-Reed-4/publication/237639208_Practical_modelling_of_nonlinear_audio_systems_using_the_Volterra_series/links/53f5ee390cf2888a7491f980/Practical-modelling-of-nonlinear-audio-systems-using-the-Volterra-series.pdf}{Link} & yes & 23 &  &  & & & & \\ \hline
1497336836 & 1985 & \href{https://heinonline.org/HOL/Page?handle=hein.journals/unilllr1985&div=9&g_sent=1&casa_token=ozh3l_JxcAMAAAAA:WeNk5OmpoVzMpwB_5dFav02vXQDIJVgidKLwGLA5f_7t4NNBhivuhwqlX8FeAU9y_FlAatz0SQ&collection=journals}{Link} & yes & 1+ & full & yes & & & & \\ \hline
1553022496 & 2007 & \href{https://heinonline.org/HOL/Page?handle=hein.journals/wmbrts16&div=17&g_sent=1&casa_token=Ch8313g7MQkAAAAA:e1XoAa-NtPPy_xM3ftPweHoSJAH7hmgrvPj0DGcIStyvVdD724M-lo8sFvAY_VdquOo1GTp-6A&collection=journals}{Link} & yes & 1+ & full & yes & & & & \\ \hline
1816860758 & 2010 & \href{https://d1wqtxts1xzle7.cloudfront.net/45345534/article-libre.pdf?1462367392=&response-content-disposition=inline%3B+filename%3DSerum_cytokine_levels_in_patients_with_h.pdf&Expires=1706281738&Signature=QrLdyGDCBwxXlN0D7XJ9OMTNIWHho7S0OTEdqHx2dD68XpGOOu4e8LQu7Q5ferO4~zAnYkHyJ4xDut2aX5D6l0S~Xk0uH3AJ0t4oPJQl9k5tcUFB4LdxUSf3-dQ-cKIS8kff7g7FNTbU-iCQ0y1nXLlz1M~y3xVcIbtchsi-L62-PhhlEhIxQAYls8N8Ln3P5Xd~Keif615VY3eDvjm96cZHSQZECKR7Q8-BwjO9Mpp~-oae6xsVw-X4kooecWQ23idpqAkTI-iAJAwaXgtURV54WoJECMsKeZPgmKaSSR3-CQHI1f5bvFW7Gb9bx6mn5~89SsL4LC2sj4~dg2AdBA__&Key-Pair-Id=APKAJLOHF5GGSLRBV4ZA}{Link} & no  & 28 &  &  & & & & \\ \hline
1957779105 & 2003 & \href{https://books.google.be/books?hl=en&lr=&id=toDBAgAAQBAJ&oi=fnd&pg=PR11&dq=Pheromones+and+Animal+Behaviour:+Discovering+pheromones}{Link} & yes & 1+ & partial & no & & & & \\ \hline
1976019262 & 1997 & \href{https://journals.aps.org/prl/pdf/10.1103/PhysRevLett.79.131?casa_token=WpFzvkxegvUAAAAA%3AFae47X3EDN6jSzRKl5Jsm9EcswbCPPSr_9ydiRPqWkaAfcX8gA5BZPzLCsdFUd_iQiS9FxA8uTbS214}{Link} & no & 14 &  &  & & & & \\ \hline
1978957683 & 1990 & \href{https://articles.adsabs.harvard.edu/full/seri/ApJ../0361//0000106.000.html}{Link} & yes & 46 &  &  & & & & \\ \hline
1989925682 & 2002 & \href{https://link.springer.com/content/pdf/10.1007/s002310100222.pdf}{Link} & no & 11 &  &  & & & & \\ \hline
2016946988 & 1967 & \href{https://rnd.edpsciences.org/articles/rnd/pdf/1967/01/ABABB_0003-388X_1967_7_1_ART0004.pdf}{Link} & yes & 17 &  &  & & & & \\ \hline
2019380115 & 1998 & \href{https://pdf.sciencedirectassets.com/271732/1-s2.0-S0278434300X00477/1-s2.0-S0278434398000326/main.pdf?X-Amz-Security-Token=IQoJb3JpZ2luX2VjEO7%2F%2F%2F%2F%2F%2F%2F%2F%2F%2FwEaCXVzLWVhc3QtMSJHMEUCIQCLxfu4fHaxyB62hivNEXQ3oZ%2BxJUNU7Ek4BZfJ3nMWogIgOyFAekrHMJkK%2BN1iPcF0FOiurS%2Bf7Dfj1tvvsVRu7d0qvAUIp%2F%2F%2F%2F%2F%2F%2F%2F%2F%2F%2FARAFGgwwNTkwMDM1NDY4NjUiDDhiY50s1P%2BHSjrY6iqQBXAb9D%2Fl3MwKk03zDnKDZFZpJo2vmLYIdVQS0gq44qnbf7OcRuq8NN2Xq5%2BiOfvnOzFJKR%2BNjGdKcffwQPuZND0%2BDZ6MMVX3w6WzRXYm%2F39dYEH5t9RNLSeuuntDMJjHLZko8ICGph1ePwnoa19RwYA2qu%2B9zfT27grMPH55OVK28ZArcpzNnghPj6EkKaYrbYmEpH6j0sWnuL0Dv58QEcAzJy1bwBfIWNRwam23CWExhKpF7gHqhdalPlzyI%2FP1zSXFDqdFpdos4OLrujQlSaf4uInw2HrAmOkgqrFK%2FLMCd9WqY0k6LXfbYnKaDA45hYIQWMIvWDLwslFPSHU0KhLDHam7opozqaCG%2F6TEao1iPF%2Brfiqpksi%2FA9AJr3Jz5NleU2QKz6HE%2BziNBosCO4Lqz1GVV%2BrlGhaHZx1KU8me%2BKMXPrHgGx77%2BCjWp2KUrgfQrY315OI4yNcGivIFT9cKcpq6CKvmDCaDDGGopvCsfFAfLcfA5%2BOxP70ABOzoCoFA26%2FwQKhl21SRhKHdw%2BjkZw%2Fd6mJJ9su4SkEdtDqvIO0vBs%2FDW2u0%2BDvJ2kbbDVKl4QE6lC%2FqVLqgoitO%2Bu3tE%2FdXwN5iC2yrD6UT0zun6pfEdplnx2E42kGBJamvSZPQslahGe5O%2F7n8aO2R7mL5bppmMqSynwXZ0iG7pwgTVKdDCTY%2BWzGc580S%2FkDiaslRRJ%2Fxs2CNHbb2%2FlqLzZvmTIibwwE7Ft5HmwOMrsJmrQofey7ozk9Lk8ymRK3lbAo1gj%2BwbUImcmAgiQ6I4qYP0Lo8DigJsJhkc0m11TcGXUQ4j8hj7u54cnkVsGcvhucFckWDDiGbe5tFDA85MaNtd9zwaxFbvI%2FT5YGuAYflMPrjzq0GOrEBeBd4L4Iq2bV9FPJUesni6AhzKReap8bLTIwkesYkDzdp49Uu%2FzH8FwfRRPyiTeBQQvlBbOc%2BSOMRfARPMvSamyRCLb%2BxeOcmMvUYinRYBFUKdVfzrIEa4InIbKiTguGM%2FSCEekkoLSa46Yg0WZh2gfFu2FUpmAVArY9XOFXeYXkyi7gBuMKJNNukTerGZhGmsneiHiV%2BQpTXvpfQtZSd7TpIP8WD9rRtpSvJkbdq4G49&X-Amz-Algorithm=AWS4-HMAC-SHA256&X-Amz-Date=20240126T143147Z&X-Amz-SignedHeaders=host&X-Amz-Expires=300&X-Amz-Credential=ASIAQ3PHCVTYVSR5E3BM%2F20240126%2Fus-east-1%2Fs3%2Faws4_request&X-Amz-Signature=8e7f80f917f9e7e681b190232b1cf9ddcd1002dc013151fa7d33d2218e0f0a32&hash=a77a7e90fd70dcc4f469df2f63963f8213eb396e7c02ab2a250d5cdb1610060c&host=68042c943591013ac2b2430a89b270f6af2c76d8dfd086a07176afe7c76c2c61&pii=S0278434398000326&tid=spdf-dd6fae97-dc8c-48bb-8103-3326b420b79e&sid=83bfcc259eb1964d4a7a0c0218a0c9b6a2a3gxrqa&type=client&tsoh=d3d3LnNjaWVuY2VkaXJlY3QuY29t&ua=02075c5a0359515152&rr=84b976282f816a91&cc=be}{Link} & no & 22 &  &  & & & & \\ \hline
2025260115 & 1998 & \href{https://www.researchgate.net/profile/David-Ikenberry/publication/253217931_Why_Active_Fund_Managers_Often_Underperform_the_SP_500_The_Impact_of_Size_and_Skewness/links/572354ca08ae262228aa6633/Why-Active-Fund-Managers-Often-Underperform-the-S-P-500-The-Impact-of-Size-and-Skewness.pdf}{Link} & yes & 16 &  &  & & & & \\ \hline
2039319856 & 1997 & \href{https://www.nature.com/articles/3300395}{Link} & no & 37 &  &  & & & & \\ \hline
\end{longtable}
\end{landscape}

\clearpage
\begin{landscape}
\begin{longtable}{|l|p{1cm}|p{4.4cm}|p{2.1cm}|p{1cm}|p{1cm}|p{1cm}|p{1cm}|p{1.2cm}|p{1.1cm}|}
\caption{Patent details.}\label{tab:patent_details} \\
\hline
\textbf{Patent ID} & \textbf{Grant Year} & \textbf{Link} & \textbf{\#References in PDF} & \textbf{US Pre-1976} & \textbf{US Post-1976} & \textbf{A1} & \textbf{E} & \textbf{Foreign} & \textbf{Other} \\
\hline
\endfirsthead

\hline
\textbf{Patent ID} & \textbf{Grant Year} & \textbf{Link} & \textbf{References in PDF} & \textbf{US Pre 1976} & \textbf{US Post 1976} & \textbf{A1} & \textbf{E} & \textbf{Foreign} & \textbf{Other} \\
\hline
\endhead

\hline
\endfoot

4295044 & 1981 & \href{https://image-ppubs.uspto.gov/dirsearch-public/print/downloadPdf/4295044 }{Link} & 5 & yes & no & no & no & no & no \\ 
\hline
5022291 & 1991 & \href{https://image-ppubs.uspto.gov/dirsearch-public/print/downloadPdf/5022291 }{Link} & 4 & yes & no & no & no & yes & no \\ 
\hline
4827022 & 1989 & \href{https://image-ppubs.uspto.gov/dirsearch-public/print/downloadPdf/4827022 }{Link} & 8 & no & no & no & no & yes & yes \\ 
\hline
7720724 & 2010 & \href{https://image-ppubs.uspto.gov/dirsearch-public/print/downloadPdf/7720724}{Link} & 11 & no & no & yes & no & yes & no \\ 
\hline

4236087 & 1980 & \href{https://image-ppubs.uspto.gov/dirsearch-public/print/downloadPdf/4236087 }{Link} & 4 & yes & no & no & no & no & no \\ 
\hline
7048812 & 2006 & \href{https://image-ppubs.uspto.gov/dirsearch-public/print/downloadPdf/7048812 }{Link} & 9 & yes & no & no & no & yes & yes \\ 
\hline
4800247 & 1989 & \href{https://image-ppubs.uspto.gov/dirsearch-public/print/downloadPdf/4800247 }{Link} & 0 & no & no & no & no & no & no \\ 
\hline
7067678 & 2006 & \href{https://image-ppubs.uspto.gov/dirsearch-public/print/downloadPdf/7067678}{Link} & 7 & no & no & yes & no & no & yes \\ 
\hline
7684359 & 2010 & \href{https://image-ppubs.uspto.gov/dirsearch-public/print/downloadPdf/7684359}{Link} & 5 & no & no & yes & no & yes & no \\ 
\hline
4230169 & 1980 & \href{https://image-ppubs.uspto.gov/dirsearch-public/print/downloadPdf/4230169}{Link} & 6 & yes & no & no & no & no & no \\ 
\hline
5596016 & 1997 & \href{https://image-ppubs.uspto.gov/dirsearch-public/print/downloadPdf/5596016}{Link} & 9 & no & no & no & no & no & yes \\ 
\hline
4331055 & 1982 & \href{https://image-ppubs.uspto.gov/dirsearch-public/print/downloadPdf/4331055}{Link} & 7 & yes & no & no & no & yes & no \\

\hline
5140549 & 1992 & \href{https://image-ppubs.uspto.gov/dirsearch-public/print/downloadPdf/5140549}{Link} & 1 & yes & no & no & no & no & no \\ 
\hline
4857516 & 1989 & \href{https://image-ppubs.uspto.gov/dirsearch-public/print/downloadPdf/4857516}{Link} & 7 & yes & no & no & no & no & yes \\ 
\hline
7304789 & 2007 & \href{https://image-ppubs.uspto.gov/dirsearch-public/print/downloadPdf/7304789}{Link} & 8 & yes & no & yes & no & yes & no \\ 
\hline
5175384 & 1992 & \href{https://image-ppubs.uspto.gov/dirsearch-public/print/downloadPdf/5175384}{Link} & 8 & no & no & no & no & no & yes \\ 
\hline
5360716 & 1994 & \href{https://image-ppubs.uspto.gov/dirsearch-public/print/downloadPdf/5360716}{Link} & 15 & no & no & no & no & yes & yes \\ 
\hline
6180695 & 2001 & \href{https://image-ppubs.uspto.gov/dirsearch-public/print/downloadPdf/6180695}{Link} & 2 & no & no & no & no & yes & no \\ 
\hline
7294680 & 2007 & \href{https://image-ppubs.uspto.gov/dirsearch-public/print/downloadPdf/7294680}{Link} & 10 & no & no & no & no & yes & yes \\ 
\hline
4359443 & 1982 & \href{https://image-ppubs.uspto.gov/dirsearch-public/print/downloadPdf/4359443}{Link} & 10 & yes & no & no & no & no & no \\ 
\hline
5065744 & 1991 & \href{https://image-ppubs.uspto.gov/dirsearch-public/print/downloadPdf/5065744}{Link} & 3 & no & no & no & no & yes & no \\ 
\hline
5568322 & 1996 & \href{https://image-ppubs.uspto.gov/dirsearch-public/print/downloadPdf/5568322}{Link} & 6 & yes & no & no & no & yes & no \\ 
\hline
4492663 & 1985 & \href{https://image-ppubs.uspto.gov/dirsearch-public/print/downloadPdf/4492663}{Link} & 9 & yes & no & no & no & no & no \\ 
\hline
\multicolumn{10}{|p{20cm}|}{\footnotesize Note: This table continues on the next three pages and contains the details of 100 randomly drawn patents from the sample with zero references and at least one forward citation, which results in $\mathrm{CD}_5=1$. The random sample is drawn from the \textit{pandas DataFrame} provided by Park et al.~\citep{park2023papers} with a random seed equal to zero.} \\

\newpage
4485566 & 1984 & \href{https://image-ppubs.uspto.gov/dirsearch-public/print/downloadPdf/4485566}{Link} & 3 & yes & no & no & no & yes & no \\ 
\hline
7592005 & 2009 & \href{https://image-ppubs.uspto.gov/dirsearch-public/print/downloadPdf/7592005}{Link} & 38 & no & no & yes & no & yes & yes \\ 
\hline

5122204 & 1992 & \href{https://image-ppubs.uspto.gov/dirsearch-public/print/downloadPdf/5122204}{Link} & 2 & no & no & no & no & yes & no \\ 
\hline
4234775 & 1980 & \href{https://image-ppubs.uspto.gov/dirsearch-public/print/downloadPdf/4234775}{Link} & 7 & yes & no & no & no & no & no \\ 
\hline
6860666 & 2005 & \href{https://image-ppubs.uspto.gov/dirsearch-public/print/downloadPdf/6860666}{Link} & 3 & yes & no & no & no & no & no \\ 

\hline
7551067 & 2009 & \href{https://image-ppubs.uspto.gov/dirsearch-public/print/downloadPdf/7551067}{Link} & 1 & no & no & no & no & yes & no \\ 
\hline
7604351 & 2009 & \href{https://image-ppubs.uspto.gov/dirsearch-public/print/downloadPdf/7604351}{Link} & 9 & no & no & yes & no & yes & yes \\ 
\hline
5342757 & 1994 & \href{https://image-ppubs.uspto.gov/dirsearch-public/print/downloadPdf/5342757}{Link} & 6 & no & no & no & no & no & yes \\ 
\hline
4865771 & 1989 & \href{https://image-ppubs.uspto.gov/dirsearch-public/print/downloadPdf/4865771}{Link} & 1 & no & no & no & no & yes & no \\ 
\hline
6967241 & 2005 & \href{https://image-ppubs.uspto.gov/dirsearch-public/print/downloadPdf/6967241}{Link} & 9 & no & no & yes & no & yes & yes \\ 
\hline
4435552 & 1984 & \href{https://image-ppubs.uspto.gov/dirsearch-public/print/downloadPdf/4435552}{Link} & 6 & yes & no & no & no & yes & yes \\ 
\hline
4242334 & 1980 & \href{https://image-ppubs.uspto.gov/dirsearch-public/print/downloadPdf/4242334}{Link} & 6 & yes & no & no & no & no & yes \\ 
\hline
4409862 & 1983 & \href{https://image-ppubs.uspto.gov/dirsearch-public/print/downloadPdf/4409862}{Link} & 6 & yes & no & no & no & yes & no \\ 

\hline
7762264 & 2010 & \href{https://image-ppubs.uspto.gov/dirsearch-public/print/downloadPdf/7762264}{Link} & 7 & no & no & no & no & no & yes \\ 
\hline
5865057 & 1999 & \href{https://image-ppubs.uspto.gov/dirsearch-public/print/downloadPdf/5865057}{Link} & 5 & no & no & no & no & yes & no \\ 
\hline
4664811 & 1987 & \href{https://image-ppubs.uspto.gov/dirsearch-public/print/downloadPdf/4664811}{Link} & 5 & yes & no & no & no & no & no \\ 
\hline
5707145 & 1998 & \href{https://image-ppubs.uspto.gov/dirsearch-public/print/downloadPdf/5707145}{Link} & 14 & yes & no & no & no & yes & no \\ 
\hline
7041814 & 2006 & \href{https://image-ppubs.uspto.gov/dirsearch-public/print/downloadPdf/7041814}{Link} & 19 & no & no & no & no & no & yes \\ 
\hline
4583720 & 1986 & \href{https://image-ppubs.uspto.gov/dirsearch-public/print/downloadPdf/4583720}{Link} & 6 & yes & no & no & no & yes & no \\ 
\hline
5581312 & 1996 & \href{https://image-ppubs.uspto.gov/dirsearch-public/print/downloadPdf/5581312}{Link} & 1 & no & no & no & no & yes & no \\ 
\hline
7502698 & 2009 & \href{https://image-ppubs.uspto.gov/dirsearch-public/print/downloadPdf/7502698}{Link} & 17 & no & no & no & no & yes & no \\ 

\hline
7659264 & 2010 & \href{https://image-ppubs.uspto.gov/dirsearch-public/print/downloadPdf/7659264}{Link} & 1 & no & no & no & no & yes & no \\ 
\hline
7805331 & 2010 & \href{https://image-ppubs.uspto.gov/dirsearch-public/print/downloadPdf/7805331}{Link} & 6 & no & no & yes & no & no & yes \\ 
\hline
4368763 & 1983 & \href{https://image-ppubs.uspto.gov/dirsearch-public/print/downloadPdf/4368763}{Link} & 5 & yes & no & no & no & yes & no \\ 
\hline
5709729 & 1998 & \href{https://image-ppubs.uspto.gov/dirsearch-public/print/downloadPdf/5709729}{Link} & 6 & no & no & no & no & yes & yes \\ 
\hline
6941475 & 2005 & \href{https://image-ppubs.uspto.gov/dirsearch-public/print/downloadPdf/6941475}{Link} & 2 & no & no & yes & no & no & no \\ 
\hline
4194052 & 1980 & \href{https://image-ppubs.uspto.gov/dirsearch-public/print/downloadPdf/4194052}{Link} & 1 & yes & no & no & no & no & no \\ 
\hline
5020086 & 1991 & \href{https://image-ppubs.uspto.gov/dirsearch-public/print/downloadPdf/5020086}{Link} & 0 & no & no & no & no & no & no \\ 
\hline
6027363 & 2000 & \href{https://image-ppubs.uspto.gov/dirsearch-public/print/downloadPdf/6027363}{Link} & 2 & yes & no & no & no & yes & no \\ 

\hline
6544700 & 2003 & \href{https://image-ppubs.uspto.gov/dirsearch-public/print/downloadPdf/6544700}{Link} & 1 & no & no & no & no & yes & no \\ 
\hline
5523292 & 1996 & \href{https://image-ppubs.uspto.gov/dirsearch-public/print/downloadPdf/5523292}{Link} & 30 & no & no & no & no & yes & yes \\ 
\hline
4573098 & 1986 & \href{https://image-ppubs.uspto.gov/dirsearch-public/print/downloadPdf/4573098}{Link} & 4 & yes & no & no & no & no & yes \\ 
\hline
4257538 & 1981 & \href{https://image-ppubs.uspto.gov/dirsearch-public/print/downloadPdf/4257538}{Link} & 5 & yes & no & no & no & no & no \\ 
\hline
4522521 & 1985 & \href{https://image-ppubs.uspto.gov/dirsearch-public/print/downloadPdf/4522521}{Link} & 13 & yes & no & no & no & no & yes \\ 
\hline
4285934 & 1981 & \href{https://image-ppubs.uspto.gov/dirsearch-public/print/downloadPdf/4285934}{Link} & 2 & no & no & no & no & yes & yes \\ 
\hline
4441704 & 1984 & \href{https://image-ppubs.uspto.gov/dirsearch-public/print/downloadPdf/4441704}{Link} & 2 & yes & no & no & no & no & no \\ 
\hline
5869693 & 1999 & \href{https://image-ppubs.uspto.gov/dirsearch-public/print/downloadPdf/5869693}{Link} & 4 & no & no & no & no & no & yes \\ 

\hline
5294907 & 1994 & \href{https://image-ppubs.uspto.gov/dirsearch-public/print/downloadPdf/5294907}{Link} & 1 & yes & no & no & no & no & no \\ 
\hline
6028059 & 2000 & \href{https://image-ppubs.uspto.gov/dirsearch-public/print/downloadPdf/6028059}{Link} & 12 & no & no & no & no & yes & yes \\ 
\hline
4353031 & 1982 & \href{https://image-ppubs.uspto.gov/dirsearch-public/print/downloadPdf/4353031}{Link} & 4 & yes & no & no & no & no & no \\ 
\hline
7633830 & 2009 & \href{https://image-ppubs.uspto.gov/dirsearch-public/print/downloadPdf/7633830}{Link} & 1 & no & no & yes & no & no & no \\ 
\hline
4238282 & 1980 & \href{https://image-ppubs.uspto.gov/dirsearch-public/print/downloadPdf/4238282}{Link} & 3 & yes & no & no & no & no & no \\ 
\hline
6403089 & 2002 & \href{https://image-ppubs.uspto.gov/dirsearch-public/print/downloadPdf/6403089}{Link} & 1 & no & no & no & no & no & yes \\ 
\hline
4383530 & 1983 & \href{https://image-ppubs.uspto.gov/dirsearch-public/print/downloadPdf/4383530}{Link} & 7 & yes & no & no & no & yes & no \\ 
\hline
4312553 & 1982 & \href{https://image-ppubs.uspto.gov/dirsearch-public/print/downloadPdf/4312553}{Link} & 3 & yes & no & no & no & yes & no \\ 

\hline
6948043 & 2005 & \href{https://image-ppubs.uspto.gov/dirsearch-public/print/downloadPdf/6948043}{Link} & 3 & no & no & yes & no & no & yes \\ 
\hline
4186183 & 1980 & \href{https://image-ppubs.uspto.gov/dirsearch-public/print/downloadPdf/4186183}{Link} & 7 & no & no & no & no & no & yes \\ 
\hline
6583828 & 2003 & \href{https://image-ppubs.uspto.gov/dirsearch-public/print/downloadPdf/6583828}{Link} & 1 & no & no & no & yes & no & no \\ 
\hline
7587403 & 2009 & \href{https://image-ppubs.uspto.gov/dirsearch-public/print/downloadPdf/7587403}{Link} & 10 & no & no & yes & no & yes & yes \\ 
\hline
7741196 & 2010 & \href{https://image-ppubs.uspto.gov/dirsearch-public/print/downloadPdf/7741196}{Link} & 12 & no & no & yes & no & no & no \\ 
\hline
4332530 & 1982 & \href{https://image-ppubs.uspto.gov/dirsearch-public/print/downloadPdf/4332530}{Link} & 4 & yes & no & no & no & no & no \\ 
\hline
4532975 & 1985 & \href{https://image-ppubs.uspto.gov/dirsearch-public/print/downloadPdf/4532975}{Link} & 5 & yes & no & no & no & yes & yes \\ 
\hline
6290363 & 2001 & \href{https://image-ppubs.uspto.gov/dirsearch-public/print/downloadPdf/6290363}{Link} & 1 & yes & no & no & no & no & no \\ 

\hline
4182612 & 1980 & \href{https://image-ppubs.uspto.gov/dirsearch-public/print/downloadPdf/4182612}{Link} & 10 & yes & no & no & no & no & yes \\ 
\hline
4331405 & 1982 & \href{https://image-ppubs.uspto.gov/dirsearch-public/print/downloadPdf/4331405}{Link} & 3 & no & no & no & no & yes & no \\ 
\hline
4269657 & 1981 & \href{https://image-ppubs.uspto.gov/dirsearch-public/print/downloadPdf/4269657}{Link} & 7 & yes & no & no & no & yes & yes \\ 
\hline
7843049 & 2010 & \href{https://image-ppubs.uspto.gov/dirsearch-public/print/downloadPdf/7843049}{Link} & 4 & no & no & no & no & yes & no \\ 
\hline
5597943 & 1997 & \href{https://image-ppubs.uspto.gov/dirsearch-public/print/downloadPdf/5597943}{Link} & 4 & yes & no & no & no & no & yes \\ 
\hline
7848211 & 2010 & \href{https://image-ppubs.uspto.gov/dirsearch-public/print/downloadPdf/7848211}{Link} & 4 & no & no & yes & no & yes & no \\ 
\hline
4656294 & 1987 & \href{https://image-ppubs.uspto.gov/dirsearch-public/print/downloadPdf/4656294}{Link} & 2 & yes & no & no & no & no & no \\ 
\hline
4587999 & 1986 & \href{https://image-ppubs.uspto.gov/dirsearch-public/print/downloadPdf/4587999}{Link} & 16 & yes & no & no & no & yes & no \\ 

\hline
5112987 & 1992 & \href{https://image-ppubs.uspto.gov/dirsearch-public/print/downloadPdf/5112987}{Link} & 3 & no & no & no & no & no & yes \\ 
\hline
5514068 & 1996 & \href{https://image-ppubs.uspto.gov/dirsearch-public/print/downloadPdf/5514068}{Link} & 4 & yes & no & no & no & no & no \\ 
\hline
7466424 & 2008 & \href{https://image-ppubs.uspto.gov/dirsearch-public/print/downloadPdf/7466424}{Link} & 8 & no & no & yes & no & no & yes \\ 
\hline
7129097 & 2006 & \href{https://image-ppubs.uspto.gov/dirsearch-public/print/downloadPdf/7129097}{Link} & 7 & no & no & yes & no & yes & yes \\ 
\hline
7030189 & 2006 & \href{https://image-ppubs.uspto.gov/dirsearch-public/print/downloadPdf/7030189}{Link} & 5 & no & no & no & no & yes & yes \\ 
\hline
5364994 & 1994 & \href{https://image-ppubs.uspto.gov/dirsearch-public/print/downloadPdf/5364994}{Link} & 2 & no & no & no & no & no & yes \\ 
\hline
4940736 & 1990 & \href{https://image-ppubs.uspto.gov/dirsearch-public/print/downloadPdf/4940736}{Link} & 2 & yes & no & no & no & no & no \\ 
\hline
5438051 & 1995 & \href{https://image-ppubs.uspto.gov/dirsearch-public/print/downloadPdf/5438051}{Link} & 4 & no & no & no & no & no & yes \\ 

\hline
6552498 & 2003 & \href{https://image-ppubs.uspto.gov/dirsearch-public/print/downloadPdf/6552498}{Link} & 1 & no & yes & no & no & no & no \\ 
\hline
4628776 & 1986 & \href{https://image-ppubs.uspto.gov/dirsearch-public/print/downloadPdf/4628776}{Link} & 6 & yes & no & no & no & yes & no \\ 
\hline
5659116 & 1997 & \href{https://image-ppubs.uspto.gov/dirsearch-public/print/downloadPdf/5659116}{Link} & 4 & no & no & no & no & no & yes \\ 
\hline
6641266 & 2003 & \href{https://image-ppubs.uspto.gov/dirsearch-public/print/downloadPdf/6641266}{Link} & 5 & yes & no & no & no & yes & no \\ 
\hline
4283043 & 1981 & \href{https://image-ppubs.uspto.gov/dirsearch-public/print/downloadPdf/4283043}{Link} & 14 & yes & no & no & no & no & no \\ 
\hline
4336212 & 1982 & \href{https://image-ppubs.uspto.gov/dirsearch-public/print/downloadPdf/4336212}{Link} & 11 & yes & no & no & no & yes & no \\ 
\hline
4438473 & 1984 & \href{https://image-ppubs.uspto.gov/dirsearch-public/print/downloadPdf/4438473}{Link} & 4 & yes & no & no & no & no & no \\ 
\hline
4364053 & 1982 & \href{https://image-ppubs.uspto.gov/dirsearch-public/print/downloadPdf/4364053}{Link} & 4 & yes & no & no & no & no & no \\

\end{longtable}
\end{landscape}

\clearpage

\end{document}